\documentclass[12pt,preprint]{aastex}
\usepackage{psfig,natbib}
\shortauthors{Bower et al.}
\shorttitle{Radio Interferometric Planet Search}
\begin{document}

\newcommand\degd{\ifmmode^{\circ}\!\!\!.\,\else$^{\circ}\!\!\!.\,$\fi}
\newcommand{\etal}{{\it et al.\ }}
\newcommand{\uv}{(u,v)}
\newcommand{\rdm}{{\rm\ rad\ m^{-2}}}
\newcommand{\msuny}{{\rm\ M_{\sun}\ y^{-1}}}
\newcommand{\mylesssim}{\stackrel{\scriptstyle <}{\scriptstyle \sim}}
\newcommand{\sci}{Science}


\title{Radio Interferometric Planet Search I:  First
Constraints on Planetary Companions for Nearby, Low-Mass Stars from
Radio Astrometry}

\author{Geoffrey C. Bower\altaffilmark{1}, Alberto Bolatto\altaffilmark{1,2}, Eric B. Ford\altaffilmark{1,3}, Paul Kalas\altaffilmark{1}}
\email{gbower@astro.berkeley.edu}

\altaffiltext{1}{Astronomy Department \& Radio Astronomy Laboratory,University of California, Berkeley, CA 94720}
\altaffiltext{2}{Department of Astronomy, University of Maryland, College Park, MD, 20742-2421}
\altaffiltext{3}{Astronomy Department, University of Florida, Gainesville, FL 32611-2055}

\begin{abstract}
Radio astrometry of nearby, low-mass stars has the potential to be a
powerful tool for the discovery and characterization of planetary
companions.  We present a Very Large Array survey of 172 active M
dwarfs at distances of less than 10 pc. Twenty nine stars were
detected with flux densities greater than 100 $\mu$Jy.  We observed 7
of these stars with the Very Long Baseline Array at milliarcsecond
resolution in three separate epochs.  With a detection threshold of
500 $\mu$Jy in images of sensitivity $1\sigma\sim100$ $\mu$Jy, we
detected three stars three times (GJ 65B, GJ896A, GJ 4247), one star
twice (GJ 285), and one star once (GJ 803).  Two stars were undetected
(GJ 412B and GJ 1224).  For the four stars detected in multiple
epochs, residuals from the optically-determined apparent motions have an
rms deviation of $\sim 0.2$ milliarcseconds, consistent with
statistical noise limits.  Combined with previous optical
astrometry, these residuals provide acceleration
upper limits that allow us to exclude planetary companions more
massive than $3-6$ $M_{Jup}$ at a distance of $\sim 1$ AU with a 99\%
confidence level.
\end{abstract}


\section{Introduction}

The study of extrasolar planets provides an important link between the
study of star formation, the study of our own solar system, and the
search for extraterrestrial life.  During the past two decades, radial
velocity surveys have over 300 extrasolar planets with $m_p \sin i
\le~13 M_J$ 
\citep[][\mbox{www.exoplanets.org}; \mbox{www.exoplanet.eu}]{2006ApJ...646..505B}.
Doppler surveys show that $\sim$10.5\% of
nearby solary-type (FGK) stars have planets with orbital periods less
than 2000 days 
\citep{2008PASP..120..531C}.

Unfortunately, several factors
make it difficult for radial velocity searches to detect planets
around M dwarfs (e.g., paucity of narrow spectral lines, reduced flux,
photospheric and chromospheric activity; 
see \citet{2006A&A...446.1165J}). From the eight M star
planetary systems detected by radial velocity techniques
\citep{2009ApJ...690..743B},
GJ 876 is particularly interesting because it has
a short-period
$\sim7.5~M_\oplus$ planet \citep{2005ApJ...634..625R},
as well as a pair of massive planets with orbital
periods of $\sim30$ and $\sim60$ days in a 2:1 mean motion resonance
\citep{2005ApJ...622.1182L}.
\citet{2002ApJ...581L.115B}
measured an
astrometric perturbation by GJ 876d of $0.25\pm0.06$mas.
GJ 849 is noteworthy because it
hosts a giant planet ($\sim~0.8 M_J$) with an orbital period of $\sim~5.1$
years 
\citep{2006PASP..118.1685B}.
\citet{2007ApJ...670..833J}
found evidence
for at least one and perhaps two giant planets orbiting GJ 317 at
separations of $\sim~$AU or more.  
Planets around GJ 849b
and GJ 317b are estimated to induce astrometric perturbations of
$\sim~0.6$mas and $\sim~0.5$mas 
\citep{2006PASP..118.1685B,2007ApJ...670..833J}.
Most recently, GJ 832 was found to have a Jupiter mass
planet in a nine-year orbit
\citep{2009ApJ...690..743B},
whereas GJ 176 has a super-Earth on a nine-day orbit
\citep{2009A&A...493..645F}.
Overall, the Jupiter-mass M dwarf planets have astrometric signals that are comparable to
those detectable with radio astrometric techniques.  

The radial velocity results allow estimates of the frequency of
short-period giant planets around M dwarfs.  
\citet{2006ApJ...649..436E}
estimate a frequency of close-in Jovian planets around M dwarfs of
$\le~1.3\%$.  
\citet{2007ApJ...670..833J} 
estimate the low-mass K an dM
stars have a $1.8\pm~1.0\%$ planet occurrence rate, significantly less
than that for solar-mass stars ($4.2\pm~0.7\%$).
Several groups are working to extend the Doppler
technique to the near-infrared 
\citep[e.g.,][]{2009IAUS..253..157L,2008PASP..120..887R,2006SPIE.6273E..72G},
though initial results will still
be biased towards short orbital periods.  Thus, it is interesting to
consider alternative techniques which are best suited for searching M
dwarfs for planets at larger separations.

The microlensing technique has discovered several extrasolar planets
around low-mass stars.  For example, 
\citep{2008Sci...319..927G}
present evidence for two giant planets with orbital separations of $\sim~2.3$
and $\sim~4.6$AU. This and two other systems 
\citep{2005ApJ...628L.109U,2008arXiv0809.2997D}
provide evidence for roughly Jupiter mass
planets. 
The host stars of the systems discovered by
microlensing are too distant to lead to a detectable astrometric
signature.  However, they do provide further evidence for a population of
planets with separations and orbital periods that are well-suited for
astrometric detection when considering nearby M dwarfs. 
Other detections 
\citep{2008ApJ...684..663B,2006Natur.439..437B,2006ApJ...644L..37G}
suggest that lower-mass planets may be quite
common around M dwarfs.

Direct imaging has been used to detect planet candidates in wide orbits
around several low-mass stars \citep{2004A&A...425L..29C, 2005A&A...438L..29C,2005A&A...435L..13N}.
%
While these planets would induce a large astrometric
perturbation, their long orbital periods would make it
impractical to discover such planets from astrometry alone.
The low density of bright M dwarfs on the sky makes transit searches
particularly difficult.  However, the potential for detecting
transiting planets near the habitable zone of an M dwarf has
motivated at least one dedicated M dwarf transit search 
\citep{2009IAUS..253...37I}.
Due to the strong bias towards finding short-period planets,
the transit technique (like the Doppler technique) is complementary to
microlensing, direct imaging, and astrometric searches that are best
suited for planets with wider orbits.

Optical astrometry has been used to search for planets around nearby
M dwarfs \citep{1996ApJ...465..264P,2003ASPC..294..107P}, leading
to discovery of several low mass companions \citep{2005ApJ...630..528P}.
These studies are fore-runners of space-based astrometric planet
searches such as with SIM \cite{2009PASP..121...41S}.

In this paper, we focus on the potential for a ground-based
astrometric planet search using techniques of radio astronomy.  Radio
astrometry has long been the gold standard for definition of celestial
reference frames \citep{2004AJ....127.3587F} and has been used to
obtain the most accurate geometric measurements of any astronomical
technique.   Astrometric results include measurement of the parallax
and proper motion of pulsars at distances greater than 1 kpc
\citep{2002ApJ...571..906B}, an upper limit to the proper motion of
Sagittarius A* of a few ${\rm\ km\ s^{-1}}$
\citep{2004ApJ...616..872R}, a $<1\%$ distance to the Taurus
star-forming cluster \citep{2007ApJ...671..546L} and to Ophiuchus
(Loinard et al. 2008), a $\sim 5\%$ distance to the Orion Nebula
star-forming cluster
\citep{2007ApJ...667.1161S,2007A&A...474..515M}, and accurate parallaxes
to star-forming regions at distances as large as 5 kpc 
\citep[e.g.,][]{2009ApJ...693..413X}.

Many radio-astrometric studies have been performed on stars. Very long
baseline interferometry (VLBI) has been used to astrometrically
discover the low-mass ($M\approx0.1$ M$_\odot$) companion of the
southern hemisphere K dwarf AB Doradus and characterize its orbit
\citep{2006ApJ...644L..37G}
Similarly, VLBI studies of radio-emitting
stars have been used to link the reference frame of the HIPPARCOS
satellite to the radio extragalactic reference frame and study the
orbit of the ternary system Algol 
\citep{1999A&A...344.1014L},
and study
the structure of the radio emission in T Tauri stars 
\citep{1996AJ....111..918P}.
In fact, T Tauri itself has been the target of a
number of astrometric studies in the radio 
\citep[e.g.,][]{2007ApJ...671..546L}.
The Very Long Baseline Array
(VLBA) can routinely achieve an astrometric accurary of $\sim100$
$\mu$as in a single epoch, and it is capable of accuracies as high as 8 $\mu$as under
favorable circumstances \citep{2003ApJ...598..704F}.   To use this
technique the target source must have a sufficiently high brightness
temperature to be detected by a high resolution radio
interferometer.  For the VLBA, brightness temperatures must be
$T_b > 10^{7}$ K, requiring nonthermal emission.  Thus, the 
systems which can be studied are limited to the most active stars.

Nonthermal stellar radio emission has been detected from many stellar
types \citep{2002ARA&A..40..217G}, including brown dwarfs
\citep{Berger06}, proto-stars \citep{2003ApJ...598.1140B}, and massive
stars with winds \citep{2005ApJ...623..447D}. Radio emission from the
late-type stars was first detected by 
\citet{1981ApJ...250..284G} and it
originates in cyclotron emission due to non-relativistic electrons in
the coronal plasma.  Only late-type stars are sufficiently bright,
numerous, and low mass to provide a large sample of stars suitable for
large-scale astrometric exoplanet searches.  Radio astrometric
searches can determine whether or not M dwarfs, the largest stellar
constituent of the Galaxy, are surrounded by planetary systems as
frequently as FGK stars and how the planet mass-period-eccentricity
relation varies with stellar type.  While Doppler and transit methods
can constrain this distributiuon at short-orbital periods, astrometry
is best suited for studying planets at a few AU, where they induce a
large astrometric signal and it is still practical to observe the
system for multiple orbital periods.   The population of gas giants at
a few AU around low mass stars is an important discriminant between
planet formation models 
\citep[e.g.,][]{2004ApJ...612L..73L,2005ApJ...626.1045I,2006ApJ...643..501B,2007Ap&SS.311....9K}.

Radio astrometric searches for planets have a number of unique
qualities.  First and foremost, searches in the radio observe stars
that because of their activity and variability are not good targets
for radial velocity or transit studies. Thus they are highly
complementary to those carried out using other techniques. Furthemore,
astrometric studies of reflex motion have the ability to fully
characterize the orbits and masses of the detected planets, without
the degeneracies inherent to radial velocity techniques.  Finally,
they are sensitive to long-period planets with sub-Jovian masses
provided there is a long-enough time baseline of observations, and
they naturally provide absolute astrometric positions tied to the
extragalactic reference frame.

The most serious limitation to astrometric accuracy may be from
stellar activity that results in an astrophysical ``jitter'' added to
the true source position.  Most evidence, however, indicates that this
jitter is small enough to permit exoplanet searches around
nearby stars.  For instance, \citet{1994ApJ...422..293W}
model the radio emission of dMe stars as originating within $\sim 1$ stellar radius
of the photosphere.  At a distance of 10 pc for a M5 dwarf, a stellar
radius is $\sim 0.1$ mas, an order of magnitude smaller than the
astrometric signature of a Jupiter analog.  The few dwarf stars
detected with VLBI appear to be compact, supporting this result
\citep{1998A&A...331..596B,2006A&A...446..733G}.

High quality astrometric positions of radio stars are also critical
for connecting radio and optical reference frames 
\citep{1997A&A...323L..49P,1999A&A...344.1014L,2007AJ....133..906B}.
Additionally, these observations will produce sizes, morphologies and
brightness temperatures critical for the study of physical
processes in active stars, which are poorly understood 
\citep[e.g.,][]{2002ARA&A..40..217G,Berger06}.

In this paper, we provide results from a flux density survey of
nearby, low mass stars (\S 2 and 3).  In \S 4, we present multi-epoch,
high-resolution astrometric observations of a subset of the detected
stars.   In \S 5, we calculate limits on companions based on the radio
measurements and archival optical astrometry.  We summarize in \S 6.

The observations described in this paper 
constitute a preliminary survey for the Radio
Interferometric Planet (RIPL) search.  RIPL is a program
with the VLBA and the 100-m Green Bank Telescope with the goal
of astrometric detection of companions to nearby, low mass stars
\citep{2007arXiv0704.0238B}.
A sample of 30 stars taken from the VLA surveys described below will
be observed 12 times over $\sim 3$ years with an astrometric
accuracy of $\sim 0.1$ mas, sufficient to detect Jupiter mass
companions at a radius of 1 AU.  The results of this paper
demonstrate that jitter in stellar position
is not a limiting factor for RIPL.

\section{Sample Definition}

Stars were drawn from two samples:  active stars in the California \&
Carnegie planet search 
\citep{2005PASP..117..657W} and X-ray selected nearby stars
which are too active for radial velocity searchs 
\citep[NEXXUS][]{2004A&A...417..651S}.
X-ray fluxes are from the ROSAT all sky survey.
The most active of the California \& Carnegie stars had
a lower X-ray luminosity than the least active of the NEXXUS
stars.  All stars are M dwarfs at $D<10$ pc.  In addition, we include
three stars, GJ436 
\citep[M2.5][]{2004ApJ...617..580B},
$\epsilon$ Eri 
\citep[K2V][]{1988ApJ...331..902C,1999ApJ...526..890C} and HD 131156 (G8V), which
are chromospherically active, nearby, and show radial velocity
signatures for planetary companions ($\epsilon$ Eri has also a nearly
face-on debris disk; 
\citet{1998ApJ...506L.133G}).
We also include the dMe star
AU Mic (GJ 803) which has a recently discovered nearly edge-on debris
disk 
\citep{2004Sci...303.1990K},
since debris disks are believed to be
signposts of planet formation.  Basic stellar data for 172 stars are
tabulated in Table~\ref{tab:stardata}. These data comprise extended Gliese
catalog number, coordinates,
error ellipse in position, proper motion, error ellipse in proper motion, 
annual parallax and its error, spectral type, apparent B and V magnitudes
and X-ray luminosity.

\section{Flux Density Observations}

We used the Very Large Array to observe our sample in June through
September 2005.  Observations were made at 5 GHz in standard continuum
mode.  The array was in B, C, and D configurations, giving a
resolution that ranged from $\sim 1$ to 10 arcsec.  Each source was
observed for 10 minutes per epoch; some sources were observed multiple
times.  Standard calibration and imaging techniques were applied with
AIPS \citep{2003ASSL..285..109G}.  The typical image rms was $\sim 50
\mu$Jy.

In Table~\ref{tab:detect} we list flux densities for sources detected
in at least one epoch.  We made 40 detections of 29 individual stars.
In Table~\ref{tab:nodetect}, we list upper limits to the flux density
for sources that were not detected.  

We plot the radio and X-ray luminosities for the detected and
non-detected stars in Figure~\ref{fig:lrlx}.  X-ray luminosities are
from the ROSAT survey and so are non-contemporaneous with  radio
observations.  Nevertheless, we see rough agreement with the
radio-X-ray luminosity correlation 
\citep{2002ARA&A..40..217G}.
We also
separate late and early type stars at type M5 but find no difference
in detection rates or flux densities.

\section{VLBA Astrometric Survey}

\subsection{Observations and Data Analysis}

In Spring, 2006, we studied the astrometric stability of seven stars
with the VLBA.  These stars were selected as among the brightest stars
from the VLA sample.  For each star, three VLBA epochs were spread
over 11 days or less (Table~\ref{tab:dates}).    Observations were
obtained at a frequency of 8.4 GHz with a recording bandwidth of 256
Mb/sec.  Phase-referenced observations were obtained for each of the
stars with an integration time on the star of $\sim 1$ hour
distributed over 4 to 8 hours, achieving an rms sensitivity of $\sim
100 {\rm\ \mu}$Jy.  Beam sizes and rms flux densities for each epoch
are in Table~\ref{tab:dates}.

Data were processed using the VLBA pipeline in AIPS.  Standard
amplitude and phase calibration techniques were employed.   The
positions and flux densities of primary and secondary calibrators are
listed in Table~\ref{tab:cals}.  In the case of primary calibrators,
the positions are those assumed for correlation.  These positions were
taken from VLBA calibrator lists and are typically accurate in an
absolute sense to a milli-arcsecond.  Secondary calibrator positions
were determined by phase referencing to the primary calibrator.  The
reported positions and flux densities are the averages over the three
epochs.

\subsection{Astrometry}

\begin{figure}[tb]
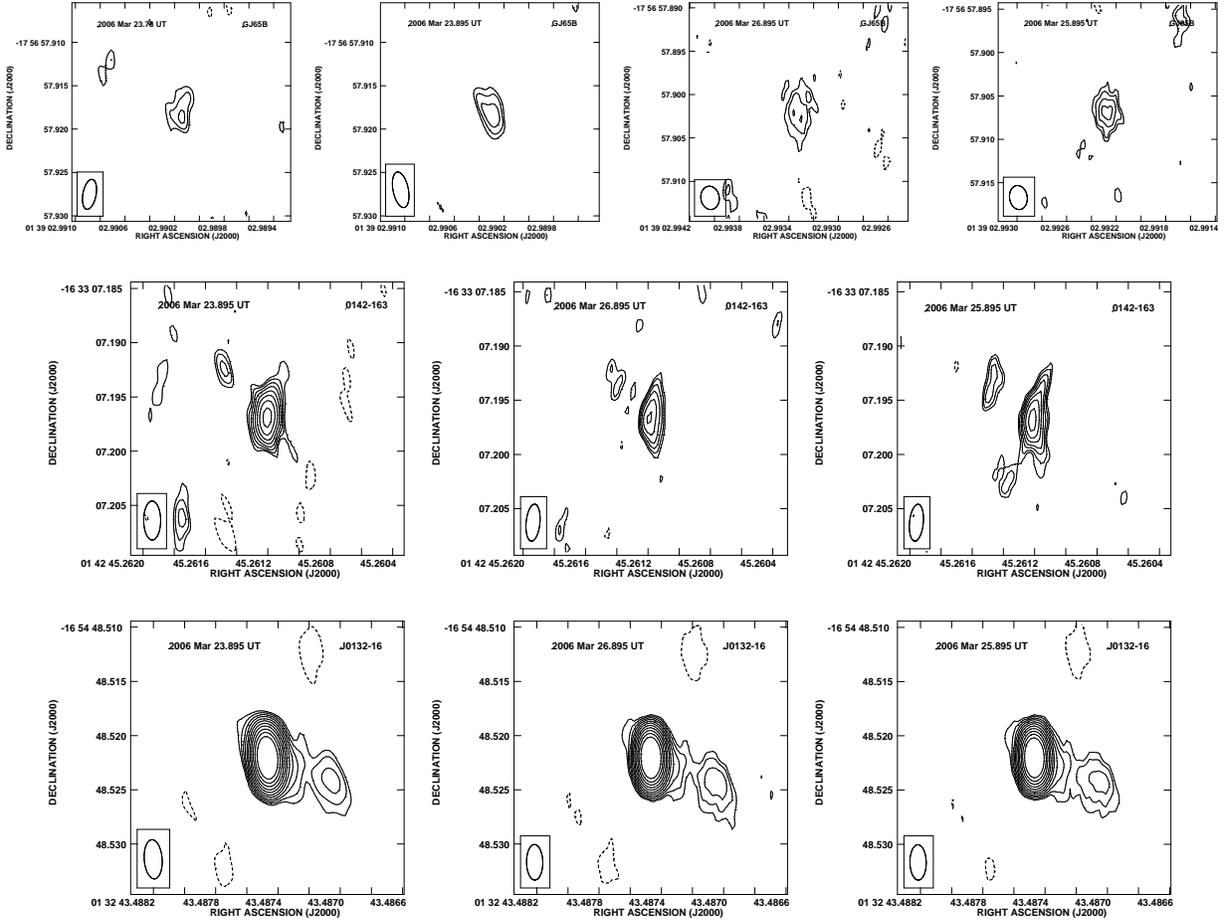

\center\mbox{
\includegraphics[width=0.2\textwidth,angle=-90]{f2a.eps}
\includegraphics[width=0.2\textwidth,angle=-90]{f2b.eps}
\includegraphics[width=0.2\textwidth,angle=-90]{f2c.eps}
\includegraphics[width=0.2\textwidth,angle=-90]{f2d.eps}
}
\center\mbox{
\includegraphics[width=0.25\textwidth,angle=-90]{f2e.eps}
\includegraphics[width=0.25\textwidth,angle=-90]{f2f.eps}
\includegraphics[width=0.25\textwidth,angle=-90]{f2g.eps}
}
\center\mbox{
\includegraphics[width=0.25\textwidth,angle=-90]{f2h.eps}
\includegraphics[width=0.25\textwidth,angle=-90]{f2i.eps}
\includegraphics[width=0.25\textwidth,angle=-90]{f2j.eps}
}
\caption[]{Images of GJ 65B and its calibrators J0142-1633 (secondary)
and J0132-16 (primary) from all
observing epochs.  Two images of GJ65B from the first and second half of the
first epoch are shown.  Contours are -3, 3, 4.2, 6, 8.4, 12, 16.8 times the
image rms noise for the star and the secondary calibrator.  
Contours are -5, 5, 7, 10, 14, ..., 160 times the image
rms noise for the primary calibrators.  The synthesized beam is shown in the
lower left.  The epoch is written in the upper left.
\label{fig:im65a}}
\end{figure}

\begin{figure}[tb]
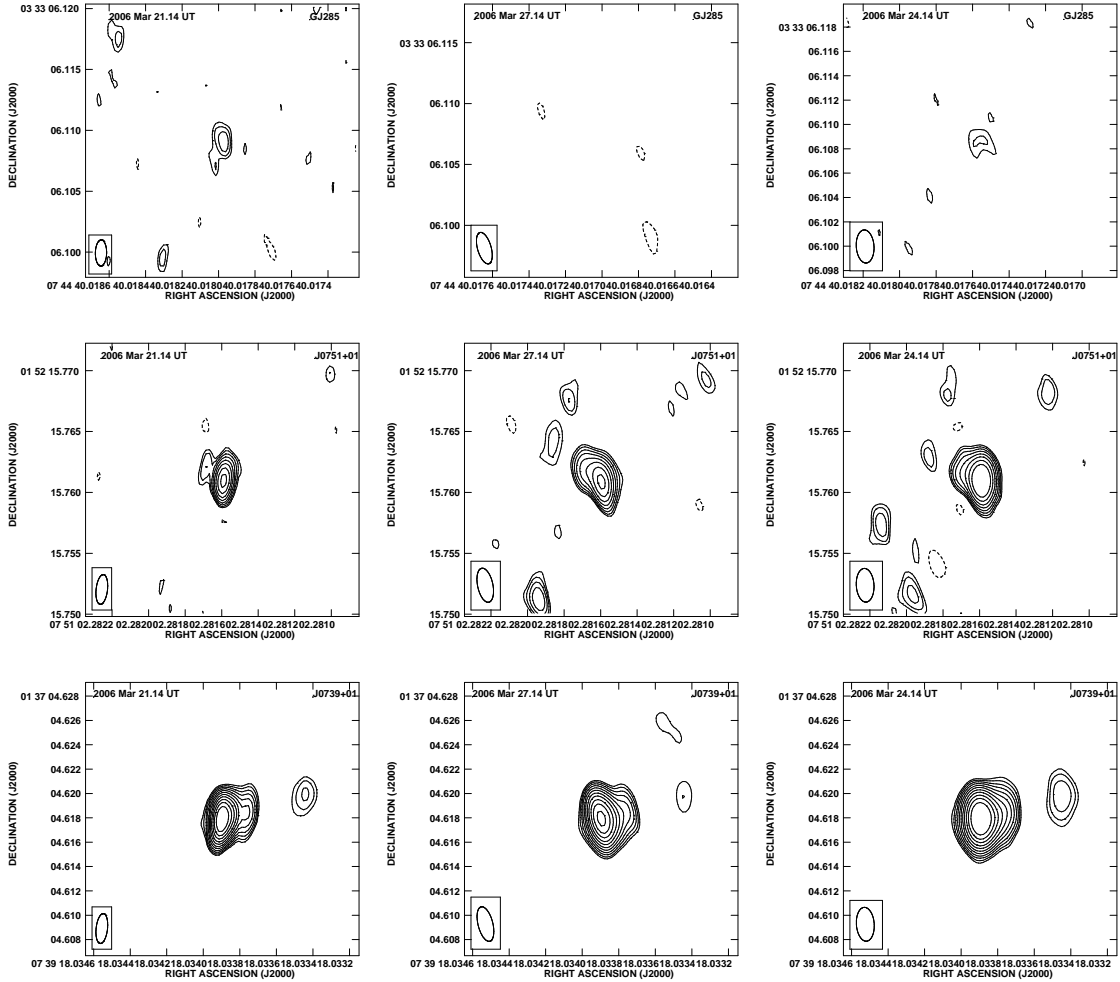

\center\mbox{
\includegraphics[width=0.25\textwidth,angle=-90]{f3a.eps}
\includegraphics[width=0.25\textwidth,angle=-90]{f3b.eps}
\includegraphics[width=0.25\textwidth,angle=-90]{f3c.eps}
}
\center\mbox{
\includegraphics[width=0.25\textwidth,angle=-90]{f3d.eps}
\includegraphics[width=0.25\textwidth,angle=-90]{f3e.eps}
\includegraphics[width=0.25\textwidth,angle=-90]{f3f.eps}
}
\center\mbox{
\includegraphics[width=0.25\textwidth,angle=-90]{f3g.eps}
\includegraphics[width=0.25\textwidth,angle=-90]{f3h.eps}
\includegraphics[width=0.25\textwidth,angle=-90]{f3i.eps}
}
\caption[]{Images of GJ 285 and its calibrators J0751+152 (secondary)
and J0739+0137 (primary). Contours and labels are as described for 
Figure~\ref{fig:im285}.
\label{fig:im285}}
\end{figure}

\begin{figure}[tb]
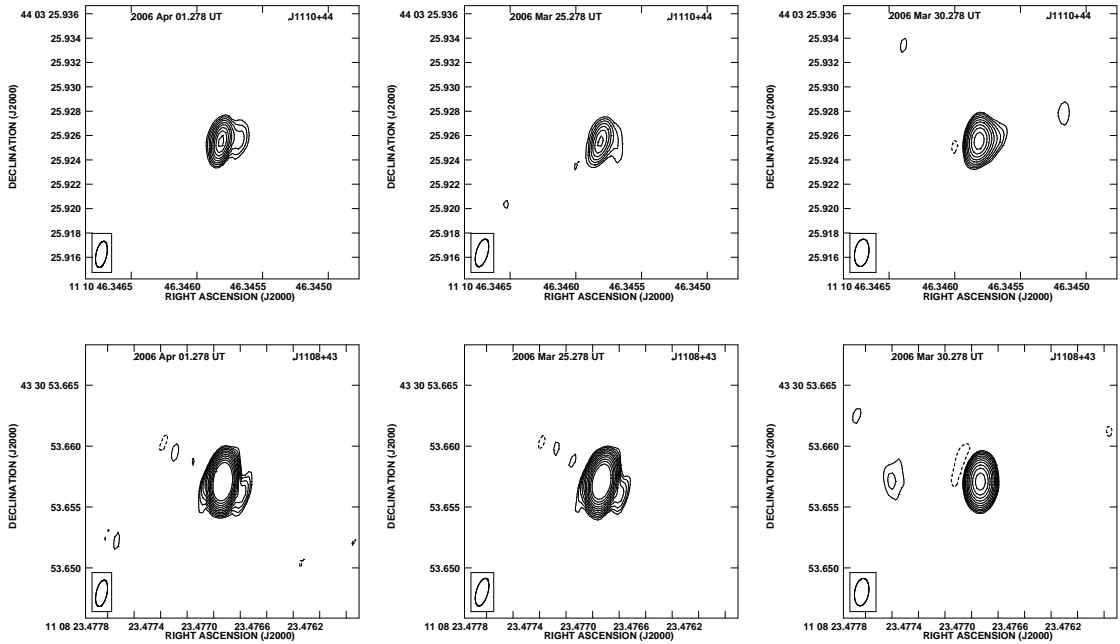

\center\mbox{
\includegraphics[width=0.25\textwidth,angle=-90]{f4a.eps}
\includegraphics[width=0.25\textwidth,angle=-90]{f4b.eps}
\includegraphics[width=0.25\textwidth,angle=-90]{f4c.eps}
}
\center\mbox{
\includegraphics[width=0.25\textwidth,angle=-90]{f4d.eps}
\includegraphics[width=0.25\textwidth,angle=-90]{f4e.eps}
\includegraphics[width=0.25\textwidth,angle=-90]{f4f.eps}
}
\caption[]{Images of the calibrators for GJ 412B:  J1110+4403 (secondary)
and J1108+4330 (primary).  Contours and labels are as described for 
Figure~\ref{fig:im65a}. 
\label{fig:im412b}}
\end{figure}

\begin{figure}[tb]
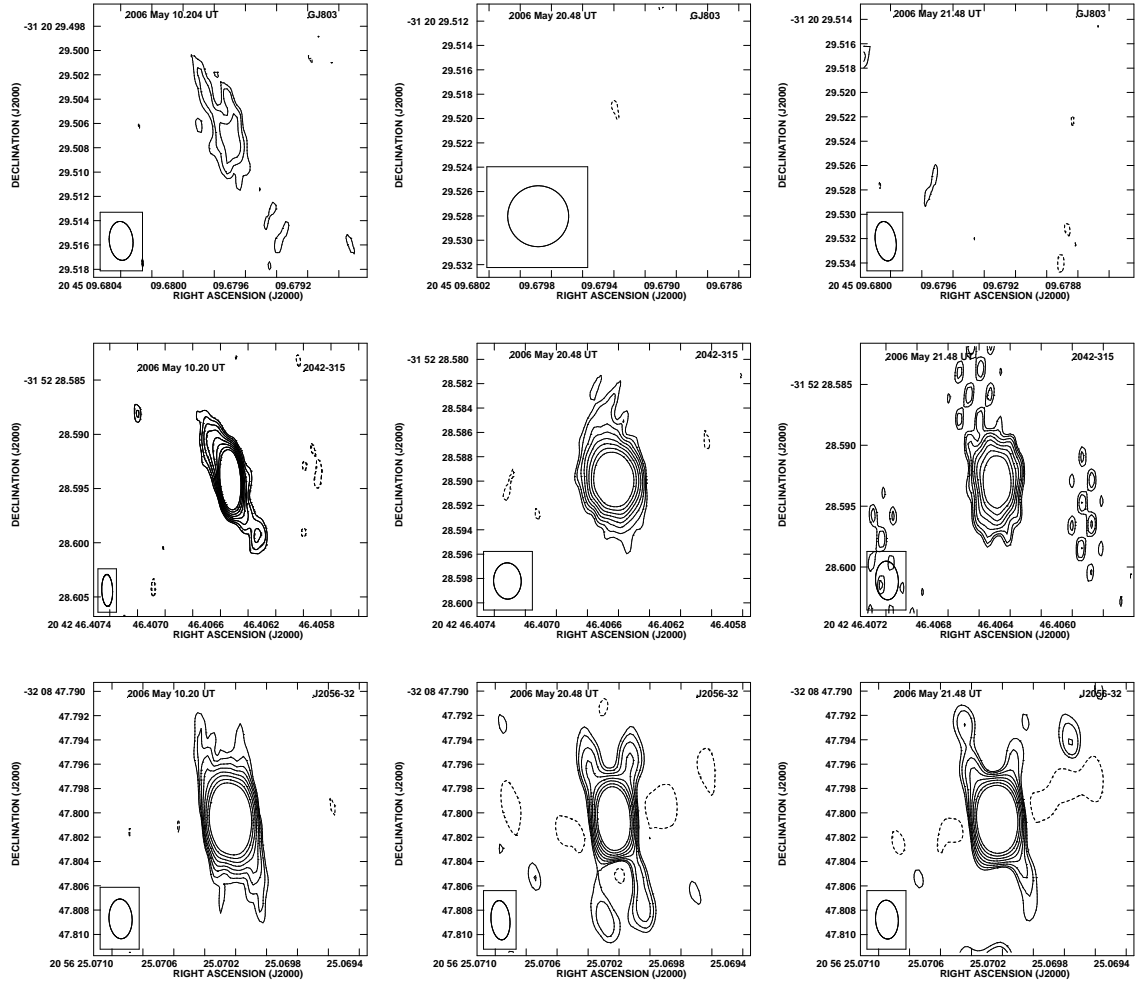

\center\mbox{
\includegraphics[width=0.25\textwidth,angle=-90]{f5a.eps}
\includegraphics[width=0.25\textwidth,angle=-90]{f5b.eps}
\includegraphics[width=0.25\textwidth,angle=-90]{f5c.eps}
}
\center\mbox{
\includegraphics[width=0.25\textwidth,angle=-90]{f5d.eps}
\includegraphics[width=0.25\textwidth,angle=-90]{f5e.eps}
\includegraphics[width=0.25\textwidth,angle=-90]{f5f.eps}
}
\center\mbox{
\includegraphics[width=0.25\textwidth,angle=-90]{f5g.eps}
\includegraphics[width=0.25\textwidth,angle=-90]{f5h.eps}
\includegraphics[width=0.25\textwidth,angle=-90]{f5i.eps}
}
\caption[]{Images of GJ 803 and its calibrators J2042-3152 (secondary) and 
J2056-3208 (primary).  Contours and labels are as described for 
Figure~\ref{fig:im65a}.\label{fig:im803}}
\end{figure}

\begin{figure}[tb]
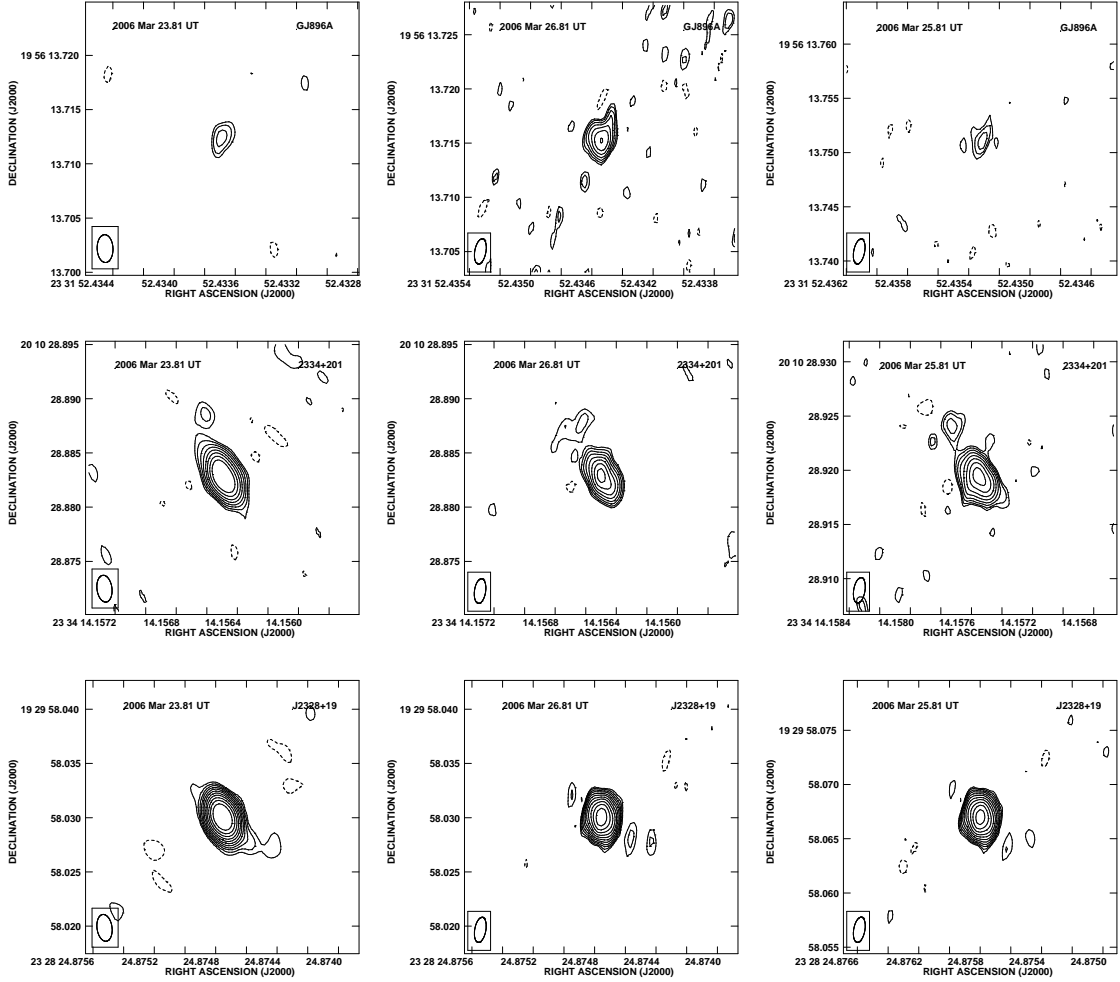

\center\mbox{
\includegraphics[width=0.25\textwidth,angle=-90]{f6a.eps}
\includegraphics[width=0.25\textwidth,angle=-90]{f6b.eps}
\includegraphics[width=0.25\textwidth,angle=-90]{f6c.eps}
}
\center\mbox{
\includegraphics[width=0.25\textwidth,angle=-90]{f6d.eps}
\includegraphics[width=0.25\textwidth,angle=-90]{f6e.eps}
\includegraphics[width=0.25\textwidth,angle=-90]{f6f.eps}
}
\center\mbox{
\includegraphics[width=0.25\textwidth,angle=-90]{f6g.eps}
\includegraphics[width=0.25\textwidth,angle=-90]{f6h.eps}
\includegraphics[width=0.25\textwidth,angle=-90]{f6i.eps}
}
\caption[]{Images of GJ 896 and its calibrators J2334+2010 (secondary) and J2328+1929 
(primary). Contours and labels are as described for 
Figure~\ref{fig:im65a}.\label{fig:im896a}}
\end{figure}

\begin{figure}[tb]
\center\mbox{
\includegraphics[width=0.25\textwidth,angle=-90]{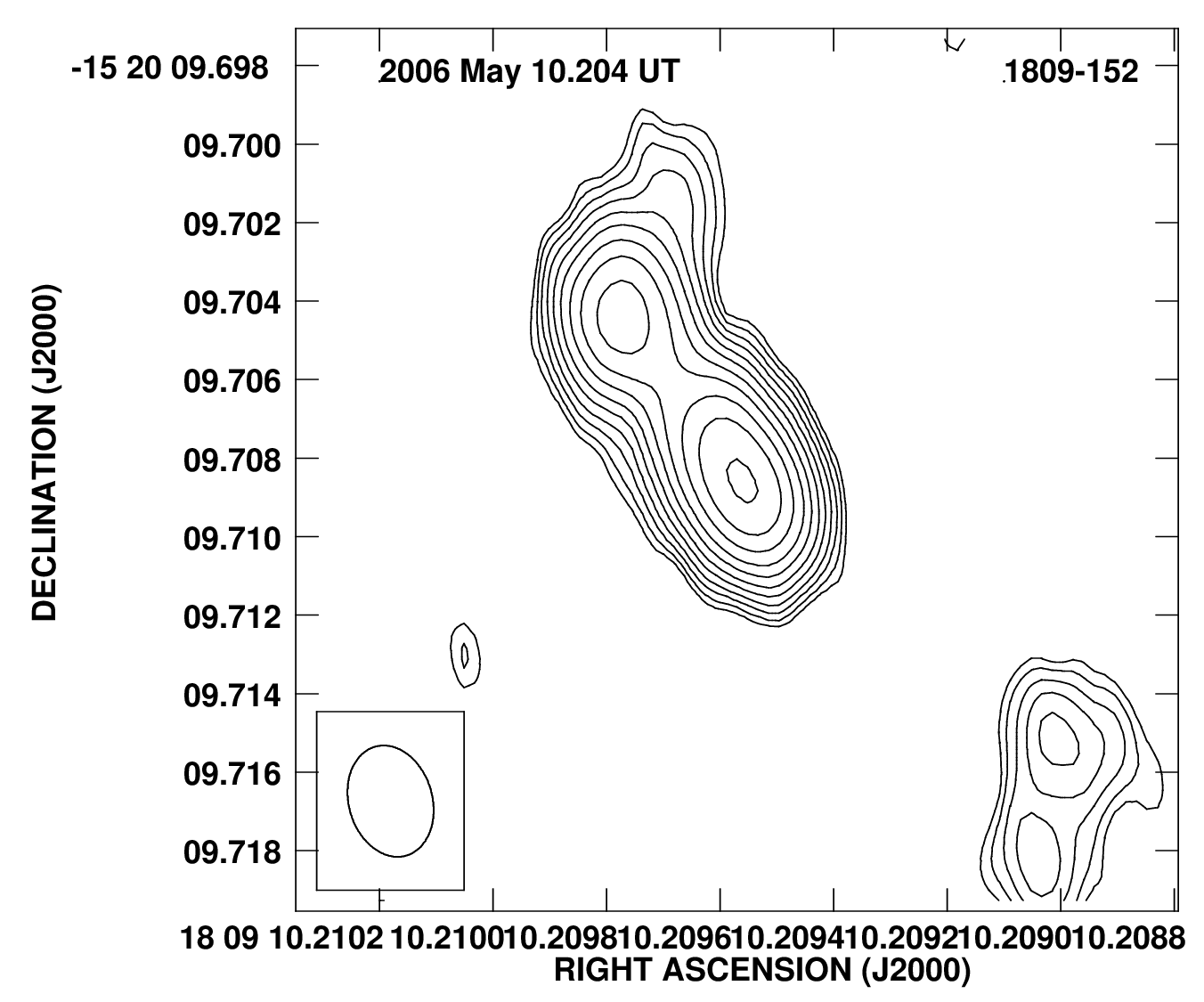}
\includegraphics[width=0.25\textwidth,angle=-90]{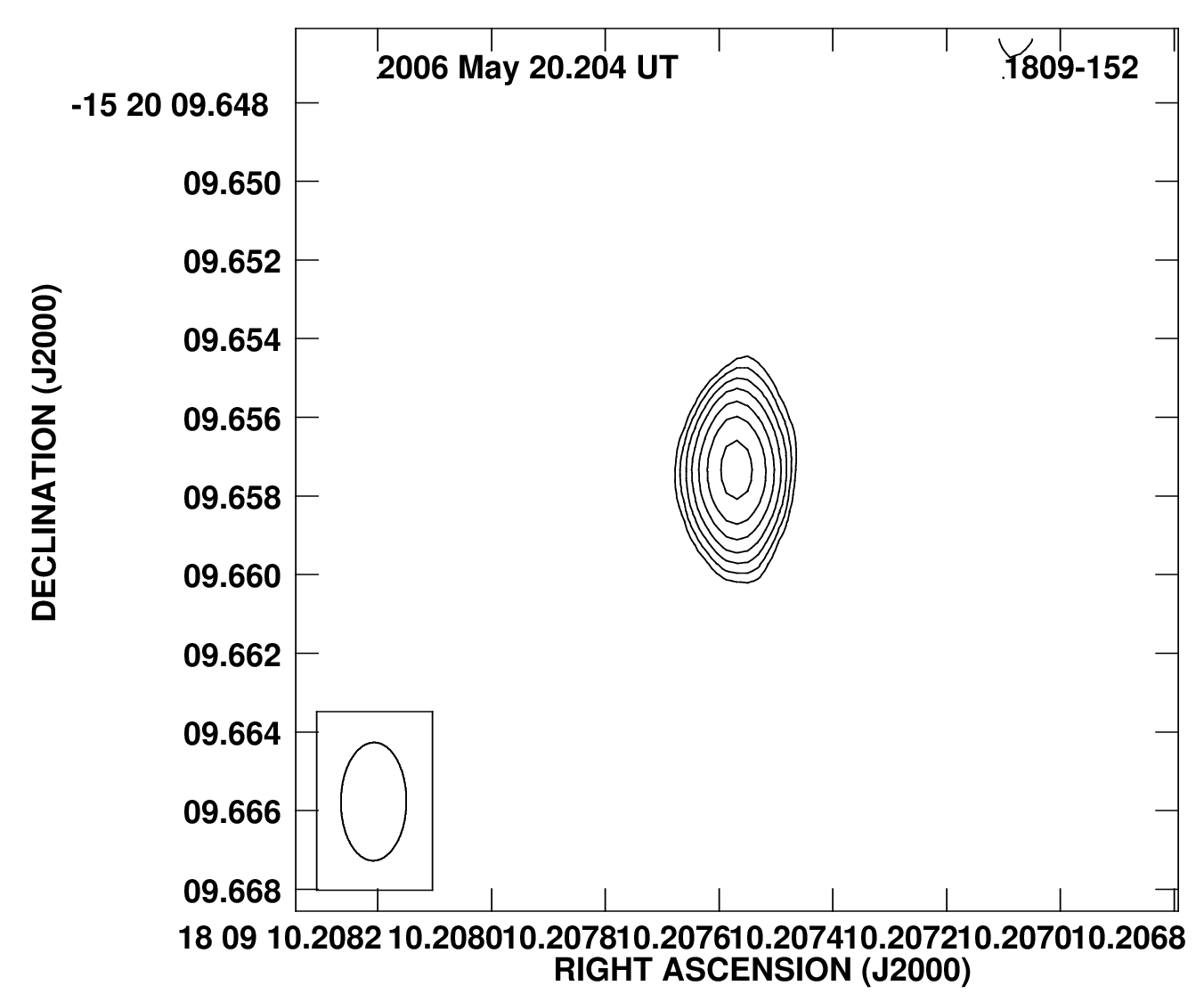}
\includegraphics[width=0.25\textwidth,angle=-90]{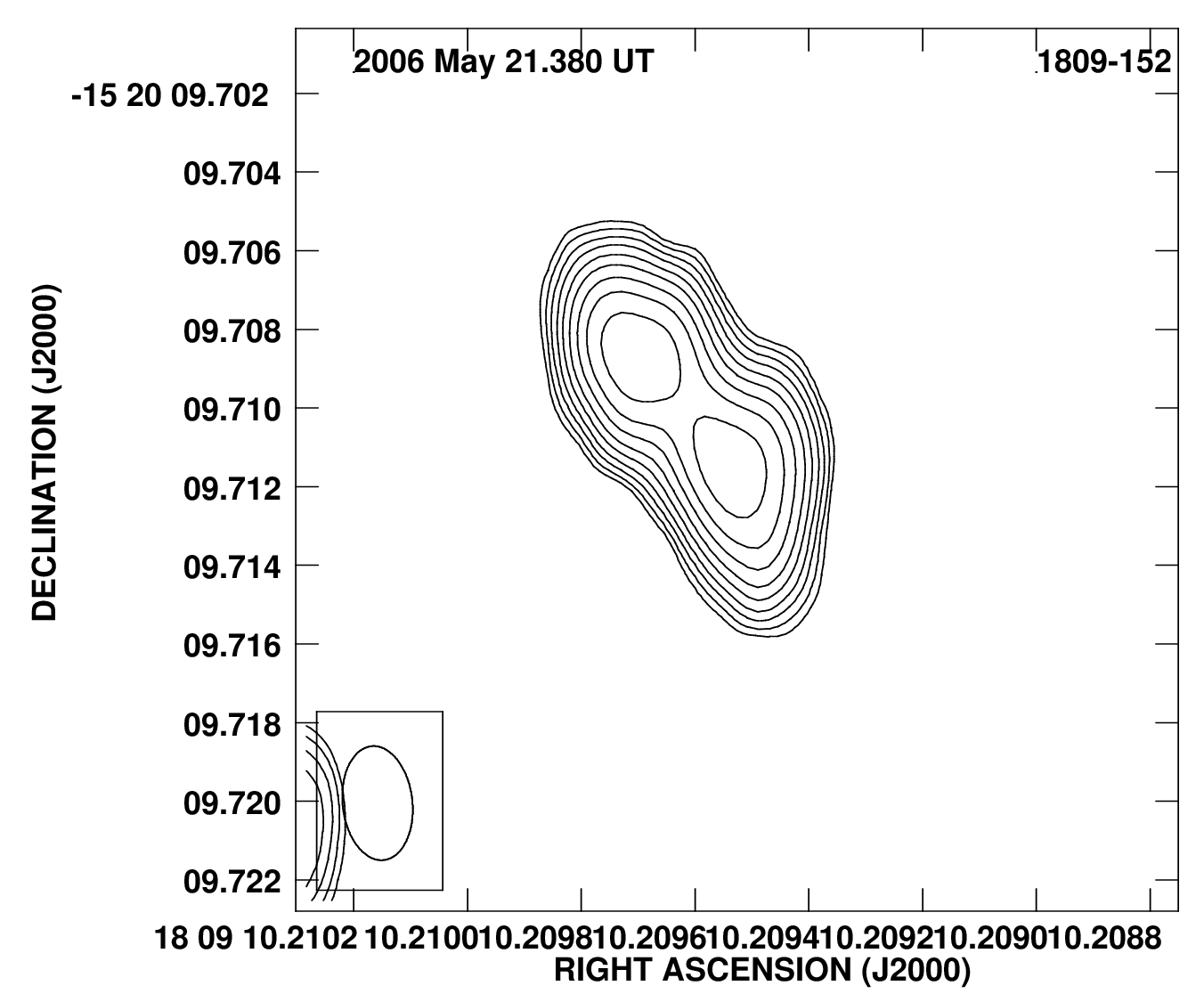}
}
\center\mbox{
\includegraphics[width=0.25\textwidth,angle=-90]{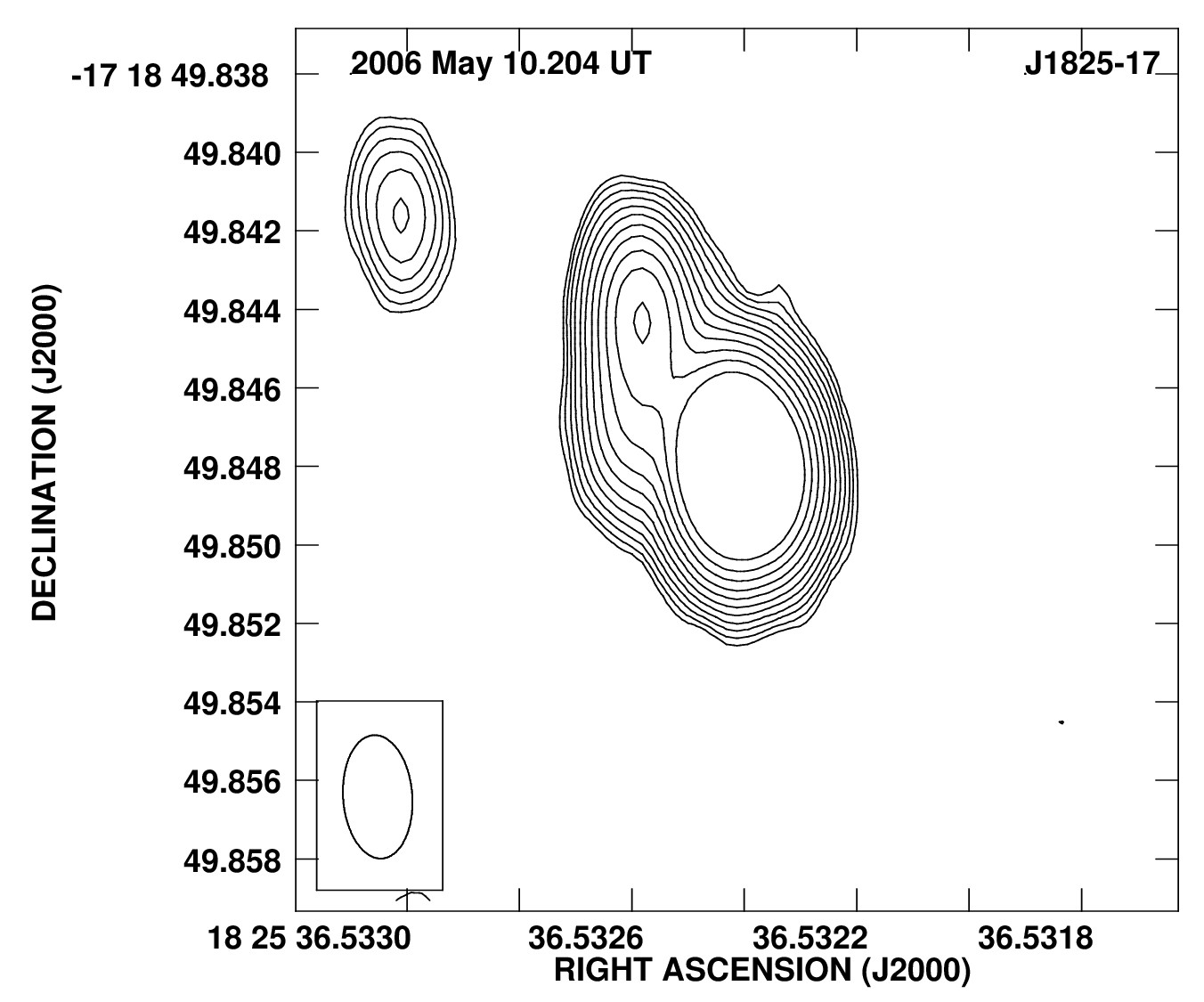}
\includegraphics[width=0.25\textwidth,angle=-90]{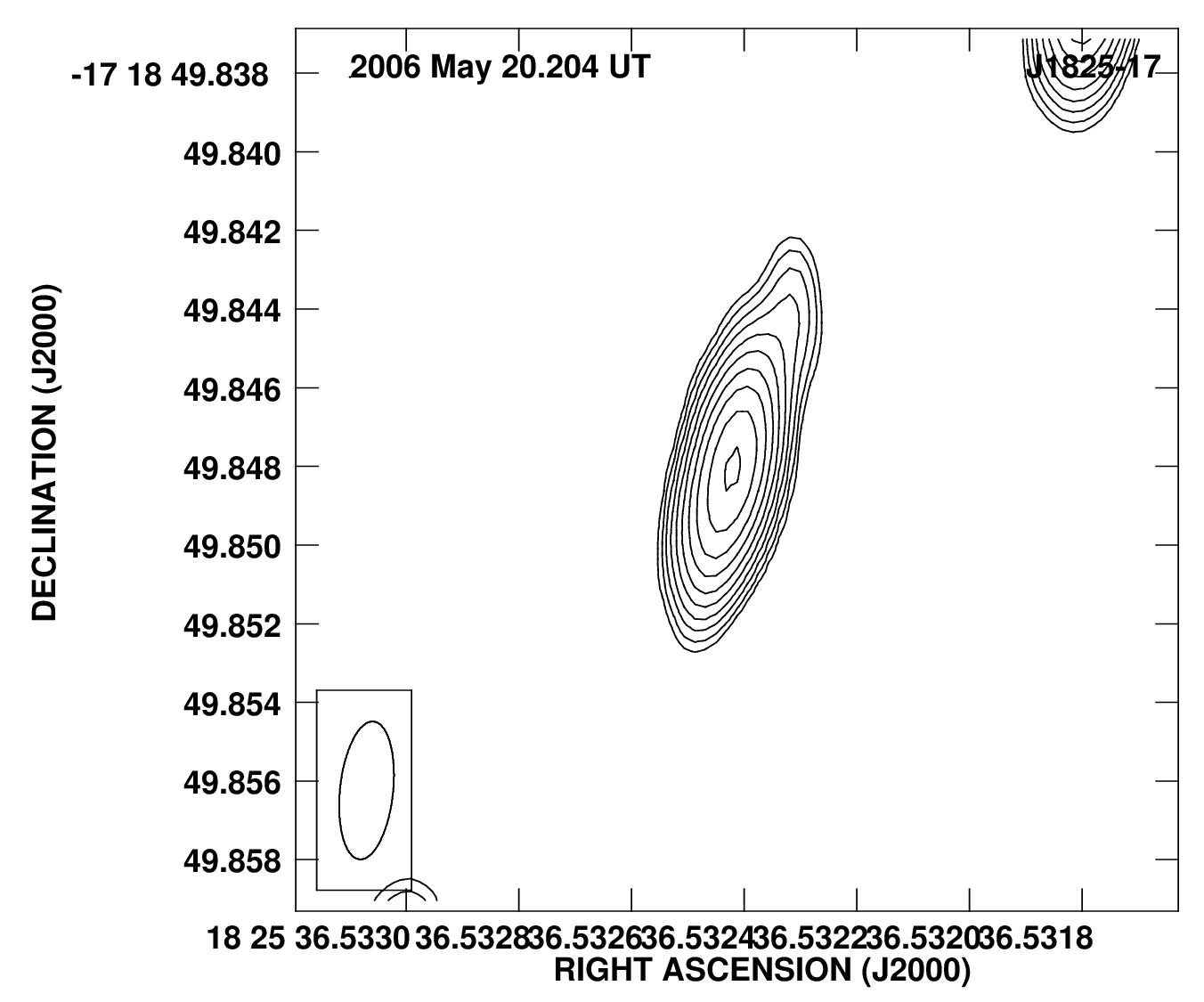}
\includegraphics[width=0.25\textwidth,angle=-90]{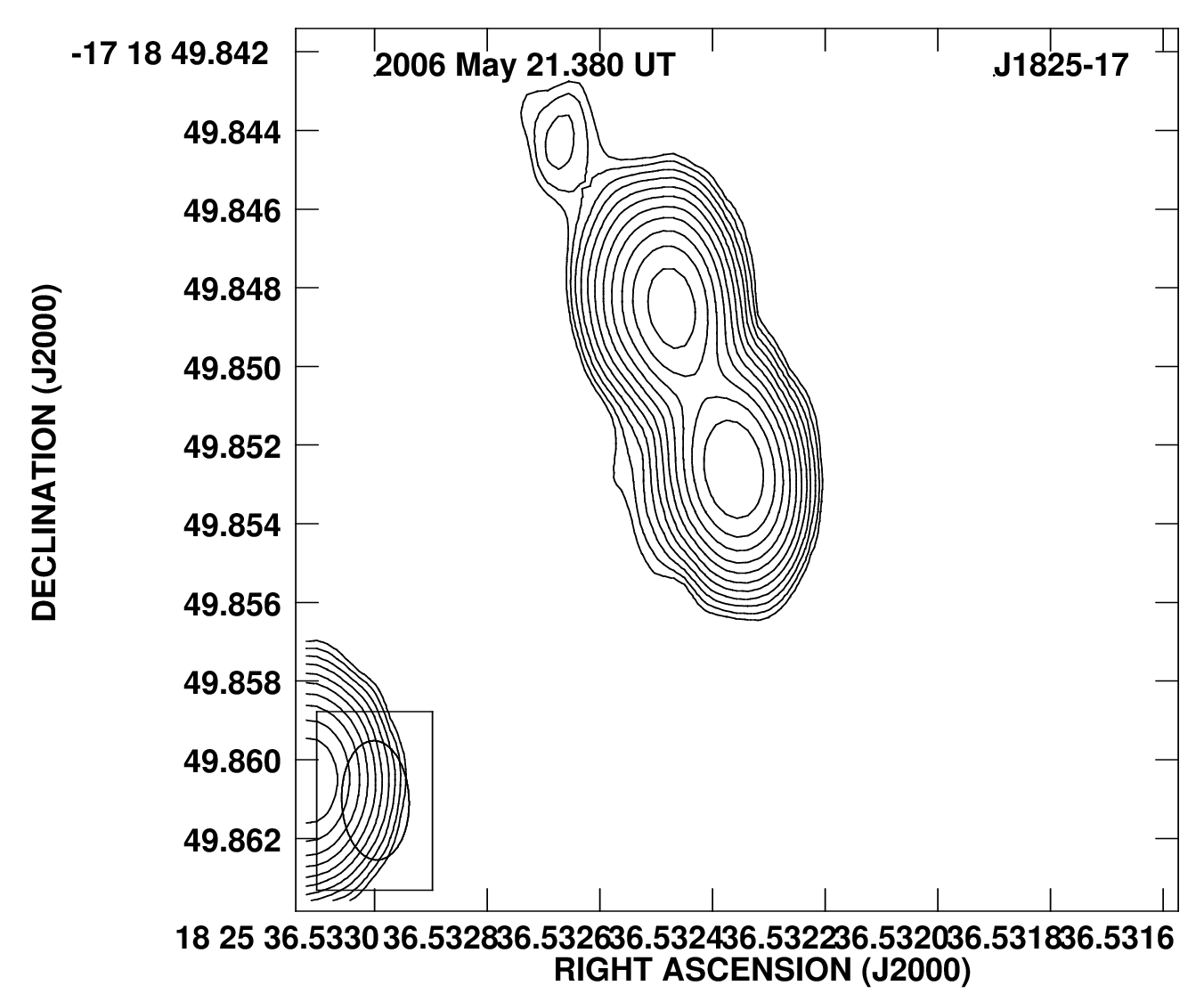}
}
\center\mbox{
\includegraphics[width=0.25\textwidth,angle=-90]{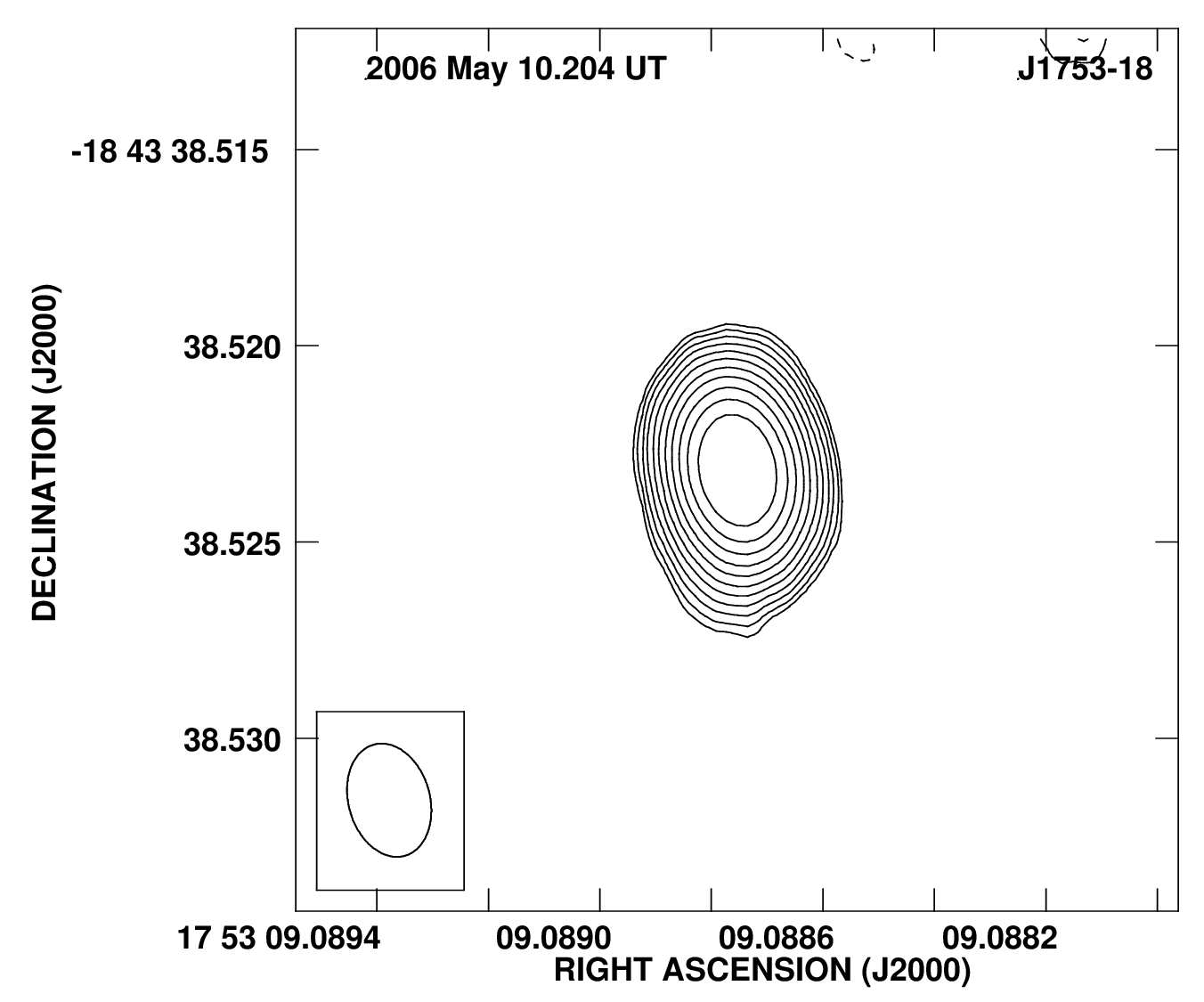}
\includegraphics[width=0.25\textwidth,angle=-90]{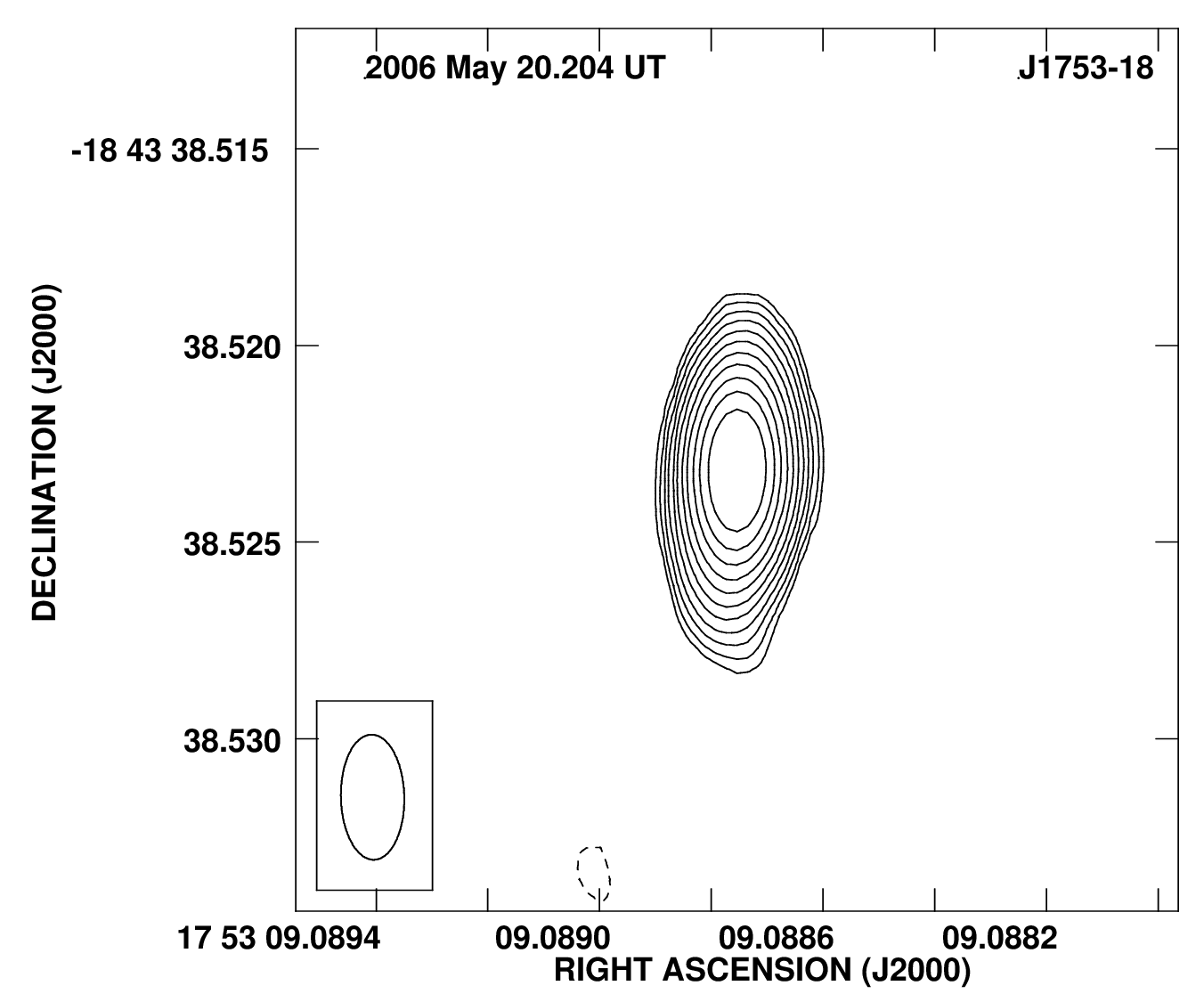}
\includegraphics[width=0.25\textwidth,angle=-90]{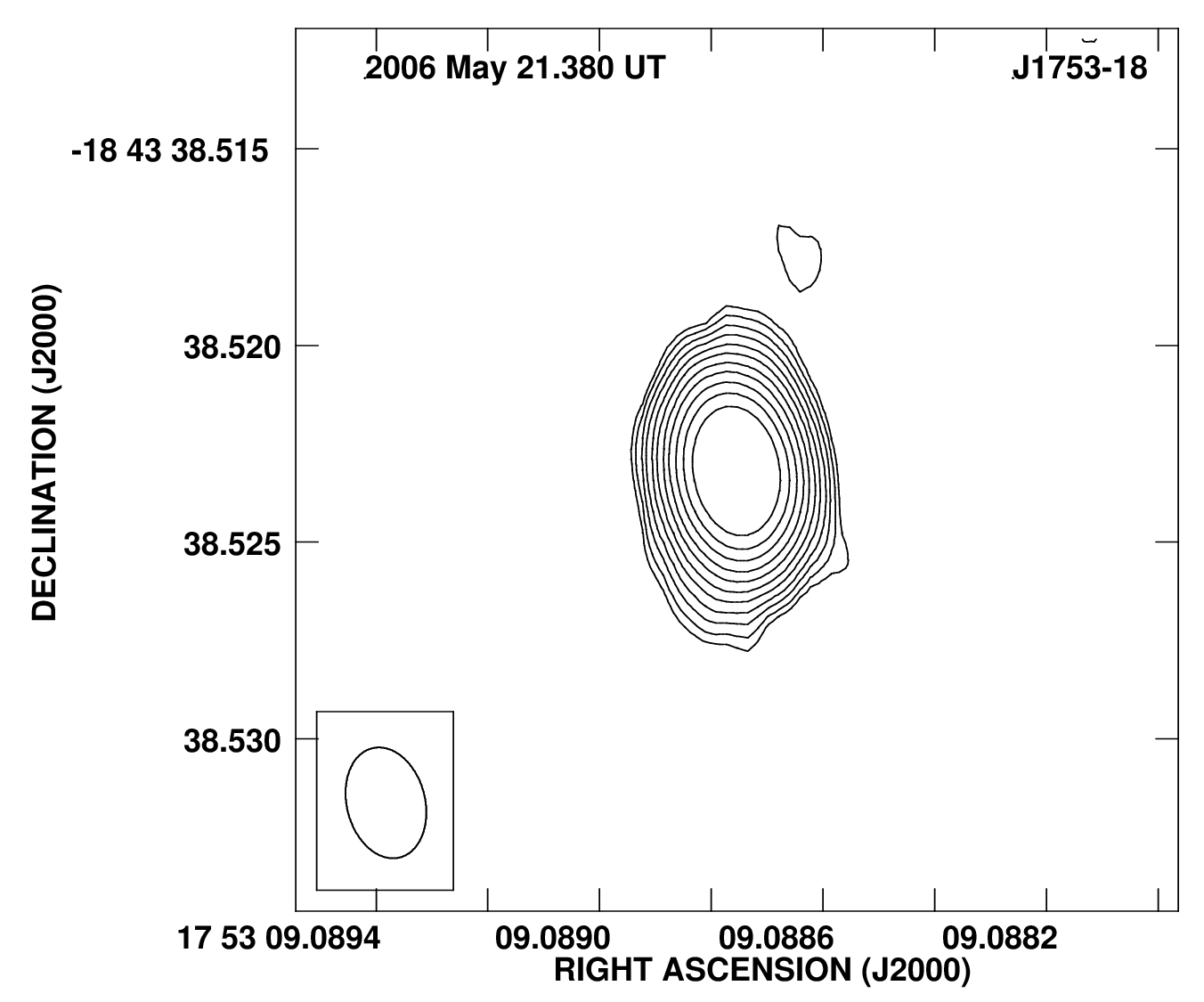}
}
\caption[]{Images of the calibrators for GJ 1224:  J1809-1520 and J1825-1718 (secondary)
and J1753-1843 (primary). Contours and labels are as described for 
Figure~\ref{fig:im65a}.\label{fig:im1224}}
\end{figure}

Of the 7 stars observed by the VLBA, three stars (GJ 4247, GJ 65B, and
GJ 896A)  were detected in all three epochs, one star (GJ 285) was
detected in two epochs, one star was detected in only one epoch (GJ
803), and two stars were not detected in any epoch (GJ 1224).  The
detections had flux densities ranging from 0.5 to 3.9 m Jy (equivalent
to SNR$\sim$ 5 to 20).  Images were reasonably well-fit as point
sources; we discuss possible deviations from point source images
later.  Best-fit positions and flux densities are given in
Table~\ref{tab:positions}.  We show images of the stars and their
primary and secondary calibrators in Figures~\ref{fig:im65a},
\ref{fig:im285}, \ref{fig:im412b}, \ref{fig:im803}, \ref{fig:im896a},
\ref{fig:im1224}.  \ref{fig:im4247},

In the second epoch of observation, GJ 285 was not detected and we
show an image at its expected position. The secondary calibrators were
imaged indicating that phase-referencing was successful.  The absence
of GJ 285 thus must be attributed to a flux density below the
$3\sigma$-limit of $\sim300\ \mu$Jy  or to resolved structure.  

GJ 65B has a large apparent motion ($\sim 16$ mas/day) and was
significantly variable during the experiment.  The apparent motion due
to both its large proper motion and large parallax causes the source
to move more than a beam width during the 5 hours of observation.   We
split epochs into two time ranges and found significantly variability
in the flux density.  In the first epoch, GJ 65B is detected
in both halves of the epoch.  In the second and third
epochs, GJ 65B is detected in only the second half of each
epoch.  Observational results are tabulated for these four half epochs
in which GJ 65B is detected.

We detected GJ 803 in only epoch.  In the second epoch, phase
stability was very poor such that the images of the calibrators were
severely distorted.  In the third epoch, the calibrator images were of
good quality but there was no detection of GJ 803 at a level of
$5\sigma$ within the field of view anticipated given the detection in
the first epoch and optical astrometry.  Imaging quality is poor for
this source due to its low declination.  For second and third epochs,
we show images at the expected locations of the star, assuming optical
proper motions and parallaxes and the position detected in the first
epoch.

Detections were not obtained in any epoch for two stars,  GJ 412B and
GJ 1224.   Observing conditions for GJ 412B were good on all three
epochs but no source was detected.  Compact calibrators J1110+44 and
J1058+43 were detected with consistent flux density, structure, and
positions in all three epochs.  GJ 412B is separated from the
calibrator by approximately $30^\prime$, which is less than the
separation of the secondary calibrators.  Thus, poor phase calibration
is unlikely to be the cause for the lack of detection of GJ
412B.  Based on optical astrometry, the apparent position of GJ 412B
was changing during these epochs at a rate $\sim 0.6$ mas h$^{-1}$,
which means that it moved more than a synthesized beam width during
the course of the observations.  If uncorrected, this leads to an
approximately 40\% reduction in the observed peak flux density.  We
corrected for this effect by making images from short time segments,
shifting the positions of these segments to account for the optical
astrometric model, and then co-adding these segments.  The star was
not detected in the final co-added images.  We searched images that
were $\sim 2.2^{\prime\prime}$ on a side, centered on the optical
position.  The likely cause of the lack of detection is low flux
density.  In two out of three VLA epochs, GJ 412B was not detected
above 200 $\mu$Jy.

In the case of GJ 1224, phase-referencing failed.  This is due to the
large offset in declination ($\sim 3$ degrees) between the phase
calibrator J1753-1843 and the star.
The image of secondary calibrator J1809-152, which is within 0.7 deg of
GJ 1224, showed the effects of poor phase calibration.  On the other
hand, the secondary calibrator J1825-1718 is within a degree in declination of
the primary calibrator and was observed to have a relatively compact
structure.  Self-calibrated observation of J1809-152 indicate a
compact source with 60 mJy flux density, suitable as a phase
calibrator for future observations of GJ 1224.

\section{Discussion}

\subsection{Survey Images}

The VLBA images in our exploratory survey, with the exception of that
for GJ65B, are predominantly compact and point-like.  This is
consistent with the radio emission originating from points close to
the stellar photosphere, which has a characteristic radius of
$R_*\approx0.1R_\sun$ \citep[equivalent to $\theta_*\approx 100$
$\mu$as at a distance of 5 pc;][]{2007ApJ...663..573B}.

The image of GJ65B (also known as UV CET B) is extended, as was previously seen by
\citet{1998A&A...331..596B}.  The image appears to consist of two
components, possibly with a variable position angle.  
\citet{1998A&A...331..596B} argue that these two components are
magnetic loops in the corona and that the star is located between
these two.  We find that the flux density of GJ65B is variable, as was
previously observed 
\citep{1983ApJ...274..776L,1985A&A...149...95P,1987ApJ...316L..85J}.

GJ 896A (EQ Peg) has been previously 
detected with VLBI and found to be compact 
\citep{1995A&A...298..187B}.
GJ 285 (YZ CMi) has been previously detected as compact
and marginally resolved with VLBI in different epochs
\citep{1991A&A...252L..19B,2000A&A...353..569P}.
GJ 803 and 4247 have not been previously imaged with VLBI
to our knowledge.
We leave a detailed investigation of stellar activity to another paper.

\subsection{Astrometric Results}

\begin{figure}[tb]
\psfig{figure=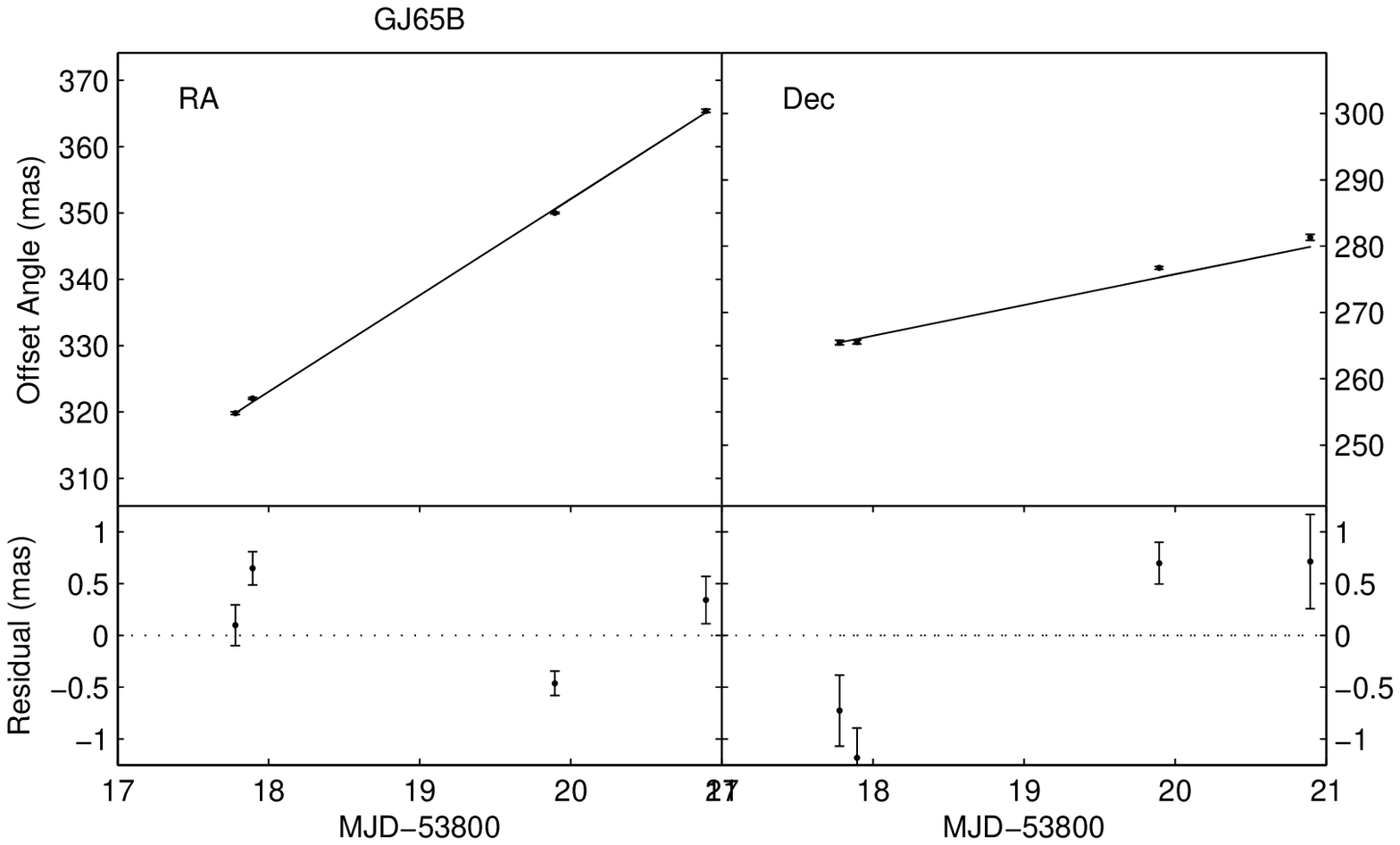}
\caption[]{Motion of GJ 65B in RA and Dec.  Plotted radio positions are relative to optical
position at first epoch.  The solid line shows the predicted optical position due
to parallax and proper motion.  The lower panels show the residual after removal of
the optical predictions and the mean offset between radio and optical.
\label{fig:pos65a}
} \end{figure}

\begin{figure}[tb]
\psfig{figure=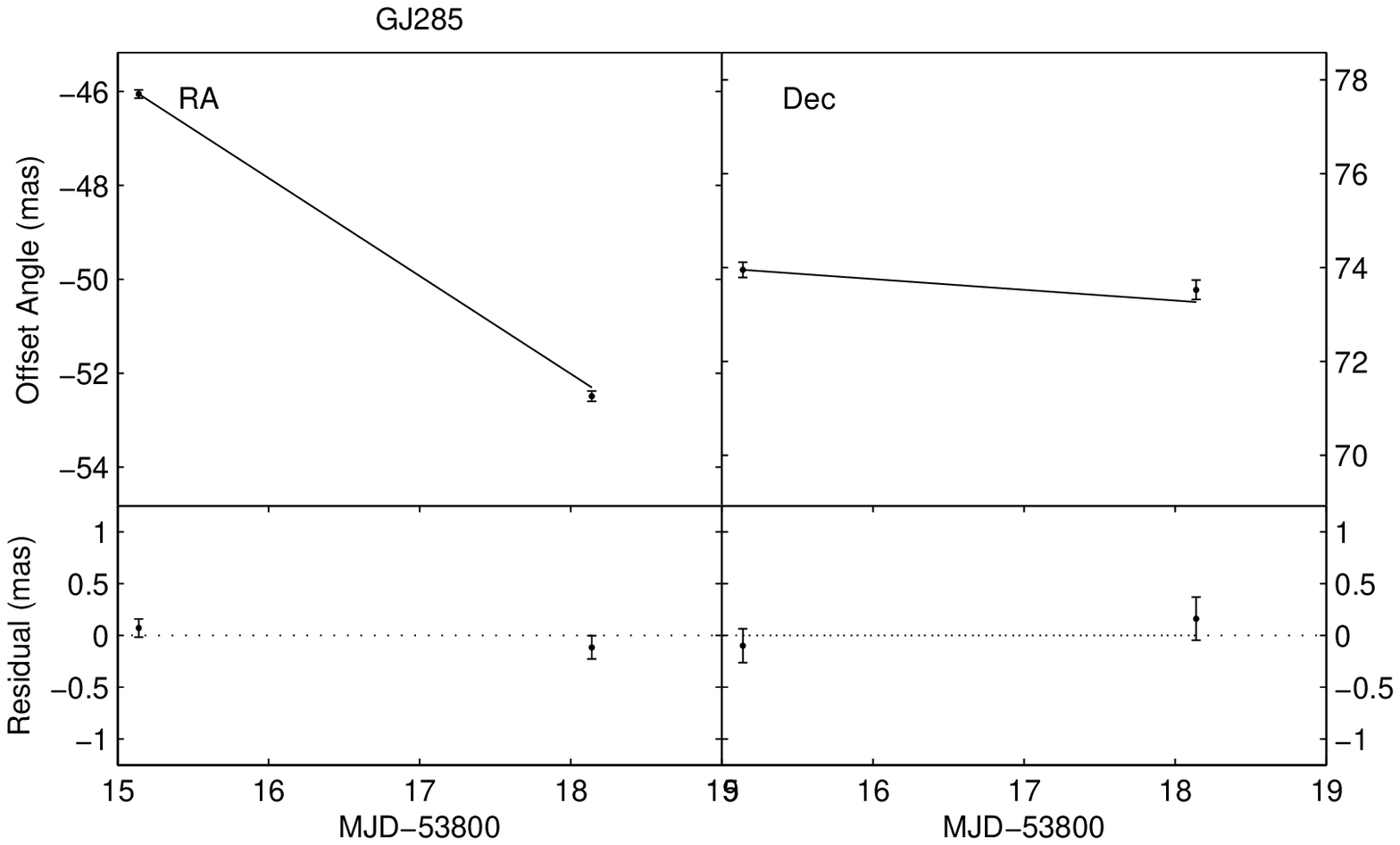}
\caption[]{Motion of GJ 285.  See Figure~\ref{fig:pos65a} for details.
\label{fig:pos285}
}
\end{figure}

\begin{figure}[tb]
\psfig{figure=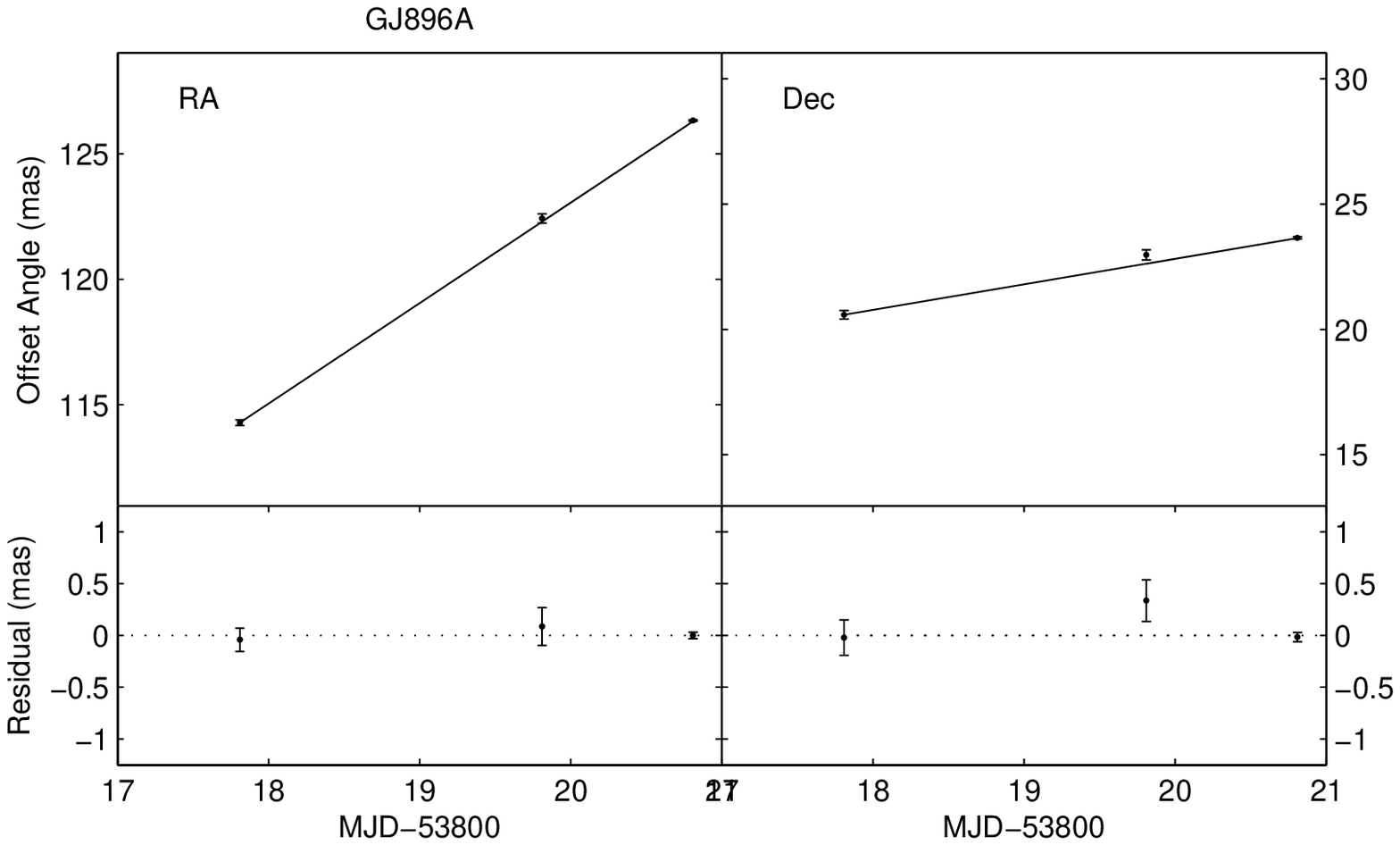}
\caption[]{Motion of GJ 896A.  See Figure~\ref{fig:pos65a} for details.
\label{fig:pos896a}
}
\end{figure}

\begin{figure}[tb]
\psfig{figure=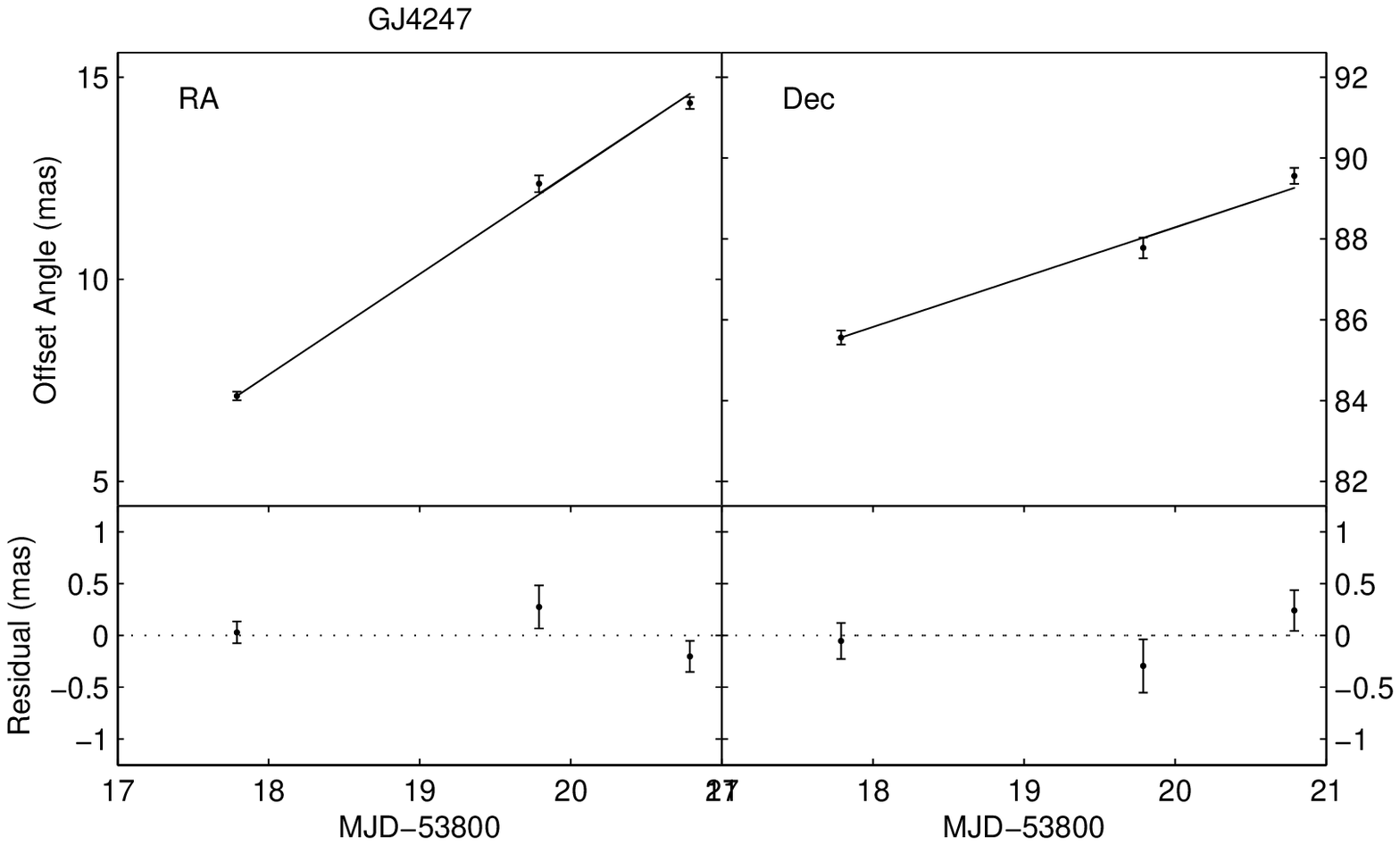}
\caption[]{Motion of GJ 4247.  See Figure~\ref{fig:pos65a} for details.
\label{fig:pos4247}
}
\end{figure}

For the four sources (GJ 896A, GJ 4247, GJ 65B, and GJ 285) detected
in multiple epochs, we plot their positions as a function of time in
Figures~\ref{fig:pos65a} through \ref{fig:pos4247}.  Positions are
referred to the expected position for the first epoch from the
existing optically-determined astrometry, proper motion, and parallax
listed in Table \ref{tab:stardata}. These originate from Hipparcos
observations for GJ 285 and GJ 896A \citep{1997A&A...323L..49P}, from
a comparison of the Luyten and 2MASS catalogs for GJ 65B
\citep{2003ApJ...582.1011S}, and from a comparison of the Luyten and
Tycho-2 catalogs for GJ 4247 \citep{2005AJ....129.1483L}.   In
Table~\ref{tab:positions}, we tabulate the radio coordinates relative
to the first epoch VLBA position ($\Delta\alpha$, $\Delta\delta$) and
to the expected optical positions ($\Delta\alpha_{\rm opt}$,
$\Delta\delta_{\rm opt}$).  

We have compared the measured positions of the radio stars with their
predicted optical positions.   The uncertainties in the radio-optical
position offsets are dominated by the relatively large astrometric
uncertainties in the optical measurements, primarily due to the
faintness of the stars and the long time baseline between the optical
and VLBA measurements.  As an ensemble, the mean positions are
consistent with no offset.  The reduced $\chi^2$ is 0.8 and
0.7 for $\alpha$ and $\delta$, respectively.  The positions
measured by the VLBA are consistent with the optical astrometry for GJ
4247, GJ 803, and GJ 285 in both coordinates.  In the case of GJ896A,
there is  a marginal detection ($3.2\sigma$) of an offset in the
absolute right ascension.  In the case of GJ65B, the total
radio-optical offset is also marginally significant. As we discuss
below, this source is part of a binary, which complicates the 
interpretation of these short observations. 

The RMS
residuals relative to the optical astrometry in each coordinate are
$\sim 0.1$ to $\sim 0.2$ mas (Figures~\ref{fig:pos65a},
\ref{fig:pos285}, \ref{fig:pos896a}, and \ref{fig:pos4247}).  
This RMS is consistent with statistical errors in the VLBA observations,
indicating that there is not a significant contribution to the source
positions from stellar activity.  The absence of a stellar activity
contribution to the astrometry is a central result of 
this paper.  It remains possible that the emission originates from 
a region that is a few stellar radii in scale but has a stable centroid on
a time-scale of days.  It is also possible that there is longer-term
stellar activity that will corrupt astrometric accuracy.

We have also compared the relative apparent motions of our sources.  
For all sources, we compute the difference in the apparent motions
determined by VLBA and that from the optical catalogs
($\Delta\mu_\alpha$ and $\Delta\mu_\delta$; Table~\ref{tab:acc}).  These
apparent motions include the effects of parallax and proper motion.  We
then compute the implied acceleration in each coordinate using the
time difference between the epoch of the VLBA observations and 
the epoch of the optical position determinations.  For
Hipparcos observations, the epoch of the optical observations is
1991.25.  For the other two catalogs, we take the epoch of proper
motion as the mean of the Luyten catalog epoch (1950) and the relevant
modern optical catalog, either 2MASS (epoch 1999) or Tycho-2 (epoch
1991.25).  The uncertainty in
acceleration has a significant component set by the relatively short time-baseline of
the VLBA measurements. For the cases of GJ 65B and GJ 4247, the VLBA observations
span only 3 days, which leads to an error in apparent motion
of $\sim 0.2 {\rm\ mas}/3 {\rm\ days} \sim 25 {\rm\ mas/year}$.  
Observations that span a year will be one to
two orders of magnitude more sensitive to acceleration.

For all stars but GJ 65B, there is no detection of an apparent
acceleration relative to the optical astrometry.  In the case of GJ
65B there is a significant offset ($6\sigma$) in the declination
proper motion.  The right ascension proper motion is fully consistent
with statistical errors in the VLBA measurements.  As we discuss
below, the apparent acceleration for GJ 65B is consistent with
the expectations for the binary orbit.

\subsection{Limits on Companion Mass and Semi-Major Axis}

For the stars for which we find an upper limit on the acceleration, we
compute limits on companion mass and semi-major axis, assuming
circular orbits.  We assume a mass of 0.1 $M_\sun$ for the star.  Our
calculation determines the fraction of systems with which a $3\sigma$
detection is made for a given set of planetary mass and semi-major
axis.

The reflex motion, $R$, of a star of mass $M$ due to an orbiting
planet of mass $M_p$ orbiting a distance $r$ away from the center of
mass of the system can be simply expressed as

\begin{equation}
R=-\frac{M_p}{M}r.
\end{equation}

Assuming a circular orbit, 
the instantaneous star speed and acceleration are

\begin{eqnarray}
\frac{dR}{dt} &\propto& \frac{M_p}{\sqrt{M r}} \\
\frac{d^2R}{dt^2} &\propto& \frac{M_p}{r^2} \label{eq:mass-acceleration}
\end{eqnarray}

\noindent where $r$ is the orbital semimajor axis of the planet and we assume
$m\ll M$. At small semi-major axes, limits on companion mass and
semi-major axis are set by the maximum angular displacement of the
star during the period covered by the observations.  At these
small separations, accelerations larger than the minimum
observable acceleration can occur, but these do not produce an
angular displacement detectable by the VLBA.
We have calculated the acceleration
of the star, projecting the systems over the full range of stellar
positions and orbital inclination angles and sum over all systems that
produce a detectable offset.  This effect then incorporates the loss
due to systems with an orbital period much less than the baseline
time.  The corresponding limit on companion masses then scales as
$r^{-1}$.

At large star-planet separations ($r$) --- and therefore long periods
--- the limits are set by the acceleration due to the companion.  The
fraction of systems that have detectable accelerations for a given
radius are a function of the inclination angle of the binary and the
position angle of the two proper motion measurements.  To determine
our confidence limits we compute the acceleration for a face-on system
and then
project the orbit onto a full range of position and inclination angles
and sum over all systems that produce an acceleration greater than our
acceleration limits.   The calculation does not take into account the
degenerate condition that occurs when the period is very close to the
separation between the optical and radio epochs.  For stars with
Hipparcos data, this peak is at 15 years, corresponding to 3 AU for
the 0.1 $M_\sun$ star; for the other stars, the time baseline is $>30$
to 50 years.   The mass limits scales with distance to the star as $r^{2}$
(see Eq. \ref{eq:mass-acceleration}).  Acceleration limits are three
times the quadrature sum of the errors in the two coordinates.

\begin{figure}
\mbox{\psfig{figure=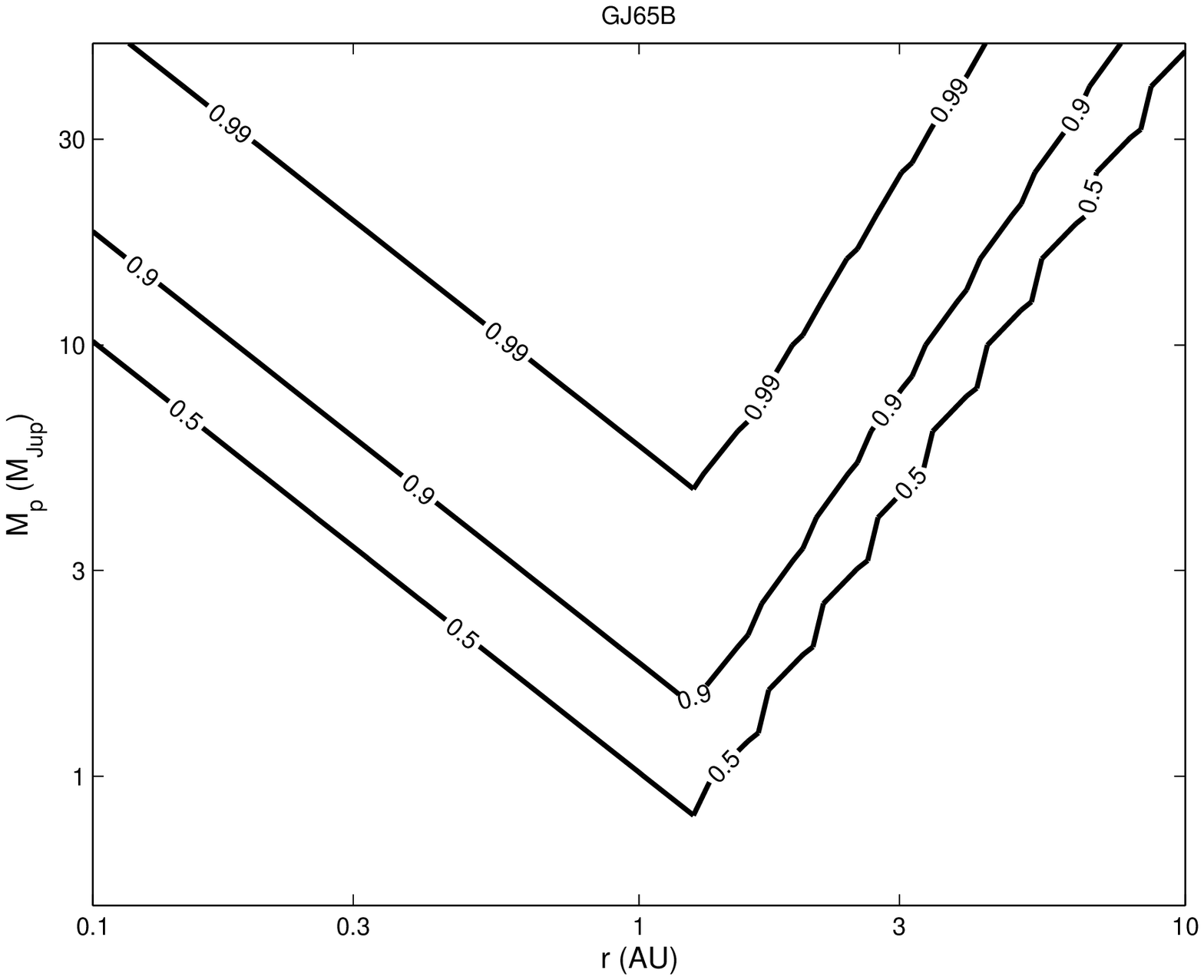,width=0.5\textwidth}\psfig{figure=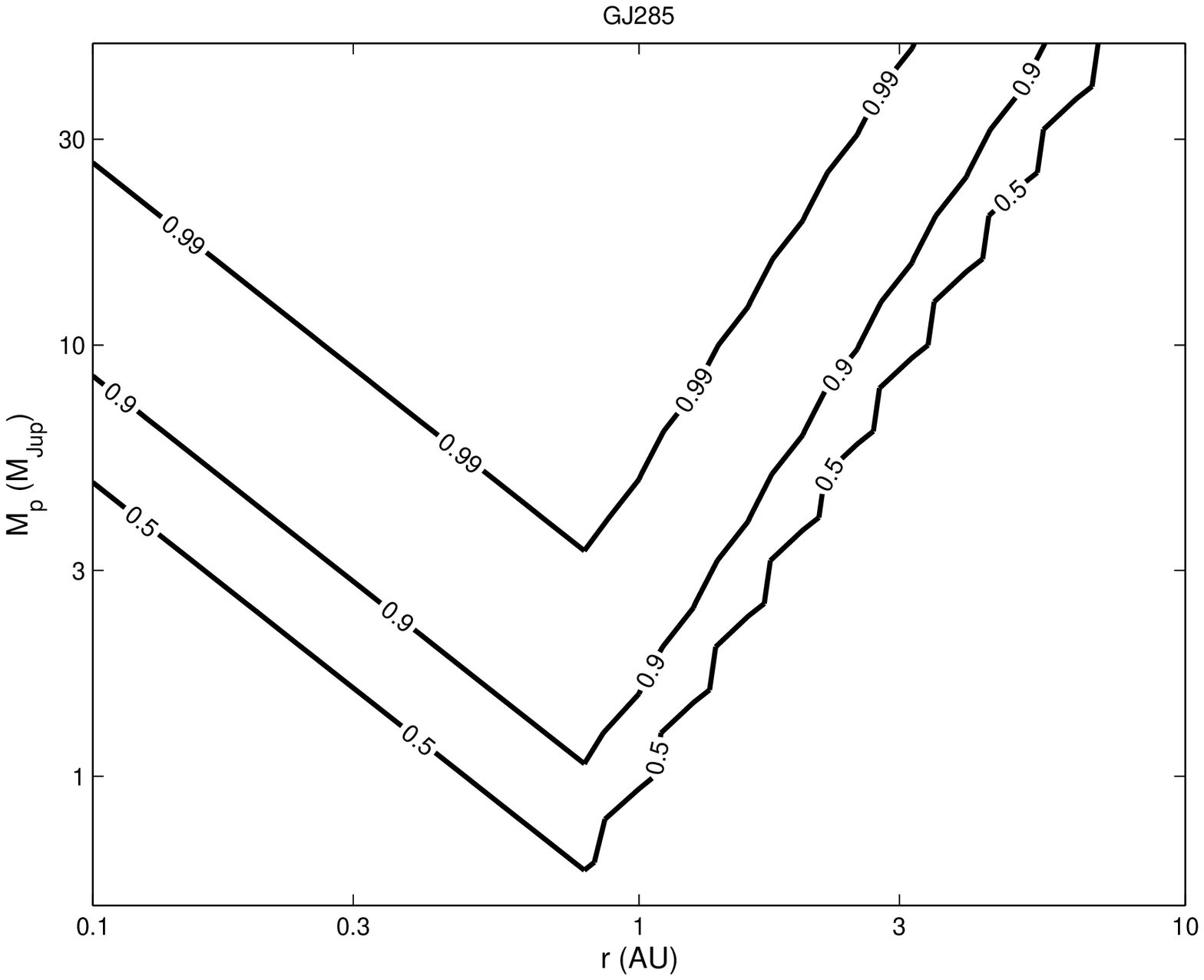,width=0.5\textwidth}}
\mbox{\psfig{figure=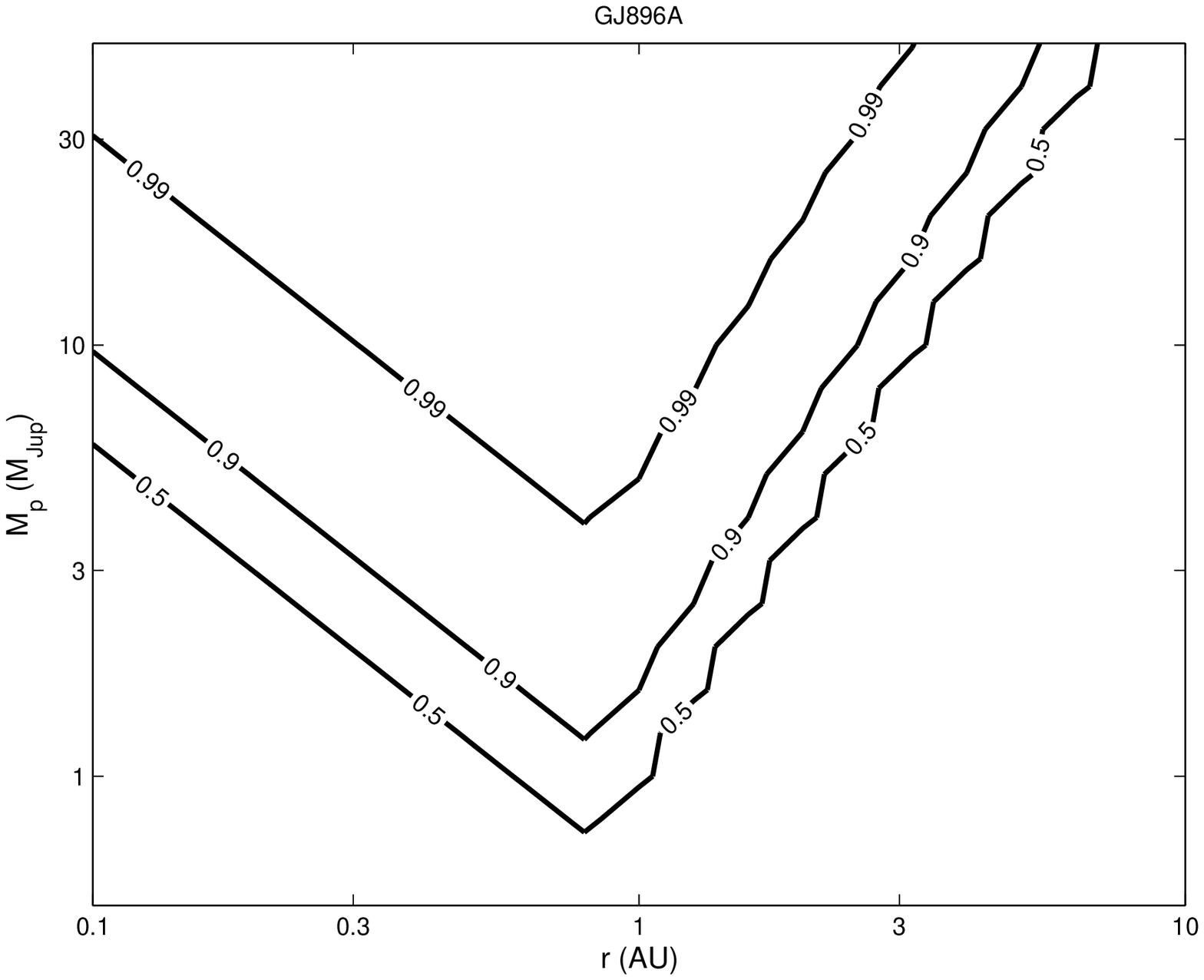,width=0.5\textwidth}\psfig{figure=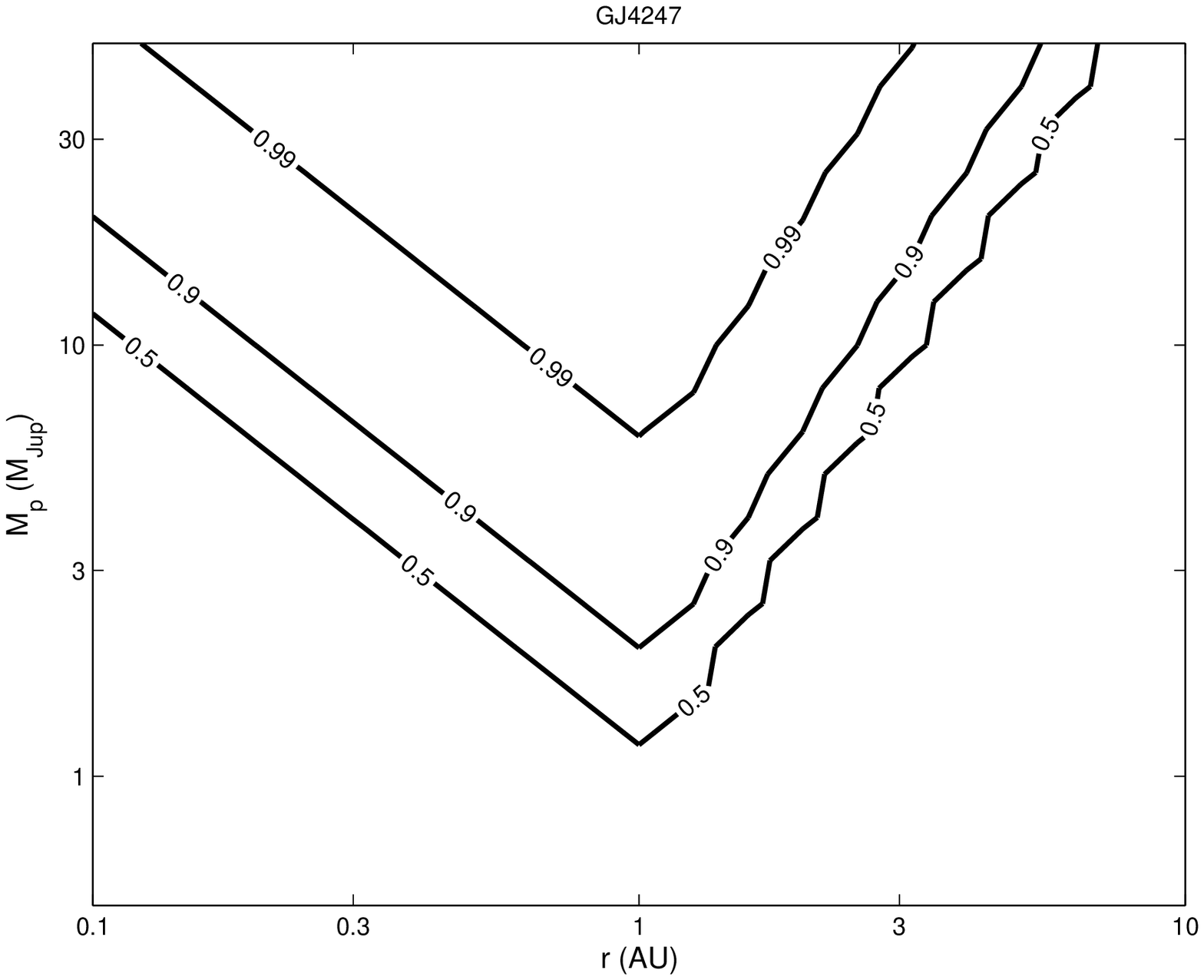,width=0.5\textwidth}}
\caption[]{Region of planetary mass and radius phase-space rejected by 
acceleration upper limits.   Contours are for parameters for which 99\%, 90\%, 50\%, 
and 10\% of systems would be detected with $3\sigma$ confidence.  Parameter
space above the curves is rejected.  Wiggles in the curves are due to  
logarithmic gridding of the model.
\label{fig:accel}}
\end{figure}

For the data presented here the acceleration limits correspond to a
minimum in mass at an approximate semi-major orbital axis $r_{min}\sim 1$ AU
(Figures~\ref{fig:accel}). That is, these observations are most
sensitive to planets orbiting at that distance from the primary. Near
this minimum the orbital periods are $\sim 3$ years and the minima in
mass ($M_{p,min}$) are in the range 2 to 5 $M_{Jup}$ (Table~\ref{tab:limits}).  We also
give the planet mass detection thresholds at 0.3 AU ($M_{p,0.3}$) and
3 AU ($M_{p,3}$); these are in the range of 10 to 20 $M_J$ and 20 to 40 $M_J$,
respectively.

To be conservative, we treat the detection of acceleration for GJ65B
as an upper limit as well.  For the nominal acceleration detection,
$a=0.0192 \pm 0.0038 {\rm AU/yr^2}$, solutions for $r > 1.4$ AU fall on the curve
\begin{eqnarray}
\left({M_p \over 4.5 M_{Jup}}\right) = \left({r \over 3 AU }\right)^2.
\end{eqnarray}
.

GJ65B has an M dwarf companion with an
orbital period $P=26.52$ yr, inclination $i=127^\circ$, and semimajor
axis $a=1.95$ arcsec or 6.3 AU \citep{1988AJ.....95.1841G}.  The
maximum apparent acceleration for this system is 0.083 ${\rm
AU/yr^2}$.  Observations at non-optimal epochs (as we have in this
case) will lead to reduced apparent acceleration, consistent with the
acceleration we have observed.  

It is difficult to model exactly the
contribution of the binary orbit to the proper motion and acceleration
of GJ65B given uncertainty over exact observing epochs and methods
for calculating the published proper motions based on
optical observations.  But it appears that the binary 
orbit was not taken into account in the proper motion calculation of 
\citet{2003ApJ...582.1011S}.  For epoch 1950 of the Luyten catalog
and epoch 2006 of the VLBA observations, the star was near 
apastron while for epoch 1999 of the 2MASS catalog the star was near
periastron.  We estimate an additional error of $(10, 30)$ mas/y in
the right ascencion and declination proper motions based on the
stellar orbit.  This has the effect of reducing the significance of
$\Delta\mu_\delta$ and the acceleration $a_\delta$ to less than $5\sigma$.
We also note that observations at low declination are prone to 
more systematic error in the declination coordinate.

GJ896A is also in a close binary system with another low mass star.
We are unaware of an orbital solution for the system and therefore
cannot place limits on acceleration due to the companion.

%


\section{Conclusions}

We measured flux densities of a sample of X-ray
selected low mass stars in the stellar neighborhood.  We detected 29
of these stars, consistent with the expectations of the radio-X-ray
correlation.  Of these stars, we observed 7 with the Very Long
Baseline Array and detected 5.  Astrometry of these stars indicates
that we are not limited by jitter in the stellar position at a level
of $\sim 0.2$ milliarcseconds.  Provided that there is not longer-term
evolution of the radio activity of these stars, our results indicate
indicate that 
radio-monitoring of these stars can be effective for detection of
Jupiter mass planets.

A comparison of radio and optical astrometry allows us to place upper
limits on companions.  We exclued companions of a $\sim 3 $ to 6 Jupiter
masses at a radius of $\sim 1$ AU for four stars.  
At radii of 0.3 and 3 AU, limits on companion masses are in the range 
of 10 to 20 $M_J$ and 20 to 40 $M_J$, respectively.
The short time-baseline of the VLBA
measurements limits the acceleration accuracy.  Longer time-scale
analysis of these sources and others will place much stricter
constraints on companion masses.    

Additionally, several
improvements to observing and analysis methods can improve
the quality of results.  Due to their large proper motion,
some of these stars change position by amounts comparable to or larger
than the size of the VLBA synthesized beam during a few hours of
observation. Use of data integrated throughout an entire
interferometric track thus results in degraded image quality and
astrometry.  We made a first-order attempt to correct this by splitting
some observations into two segments and analyzing them independently; however,
more sophisticated approaches in the visibility domain will generate 
higher quality results.  Accurate handling of known orbits
for binary stars is also necessary to achieve full sensitivity to planetary
companions.  Fortunately, known binary companions have periods that are 
long compared to the observing period and so residuals are likely to 
appear as linear terms that can be removed through proper motion fitting.
Further, we have performed a simple analysis of orbital parameters given
the limited nature of this data; future modeling must solve for orbital
parameters in a more complete and thorough manner.
Finally, greater sensitivity through increased recording
bandwidth and the use of larger apertures can increase SNR and push
astrometry accuracy to $0.1$ milliarcseconds and smaller.

This paper is the first result from the VLBA+GBT Radio Interferometric
Planetary (RIPL) survey.  Future papers will investigate the
detectability of other stars at high angular resolution, the long-term stability 
of stellar positions, and the detection of planets through astrometric means.

\acknowledgements

We thank Jason Wright for generously sharing data that 
helped define the RIPL target sample.  
We wish to thank Andrew Howard and Stephen White for their
useful comments.  The National
Radio Astronomy Observatory is a facility of the National Science
Foundation operated under cooperative agreement by Associated
Universities, Inc.  This research has made use of the NASA/IPAC
Extragalactic Database (NED) which is operated by the Jet Propulsion
Laboratory, California Institute of Technology, under contract with
the National Aeronautics and Space Administration. This research has
made use of the SIMBAD database, operated at CDS, Strasbourg, France.


\plotone{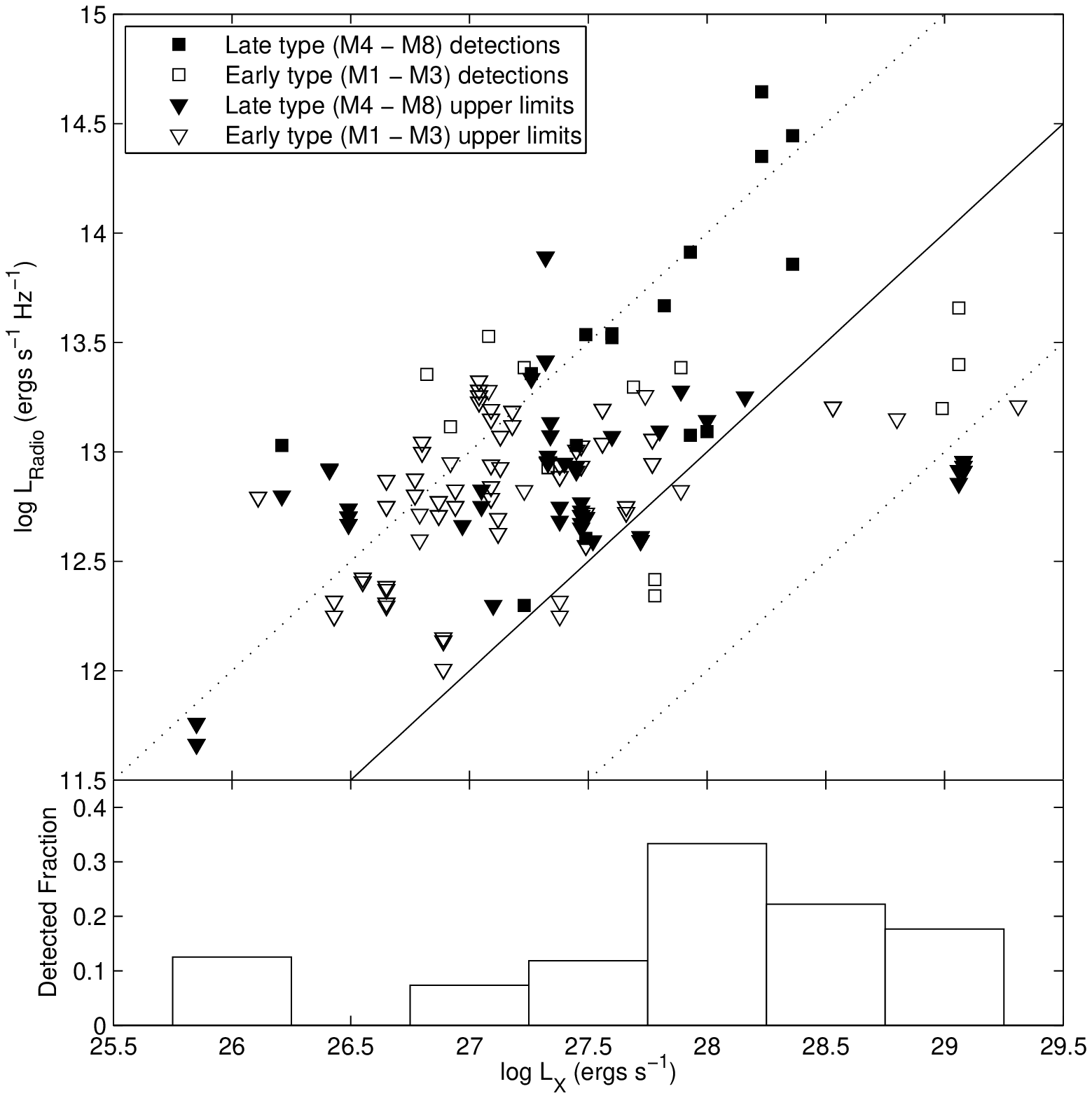}
\figcaption{Radio and X-ray luminosities.  
Results are plotted separately for late
type (filled symbols) and early type (unfilled symbols), where the dividing spectral
type is M4.  Squares represent detections and triangles
represent upper liits.  The solid line is the nominal correlation $L_R = 10^{-15} 
L_X$ Hz.  Dashed lines represent the standard uncertainty of an order of 
magnitude in the scale factor.  A histogram of detected fraction as a function
of $L_X$ is also given.
\label{fig:lrlx}}

\begin{figure}[tb]
\center\mbox{
\includegraphics[width=0.25\textwidth,angle=-90]{f8a.eps}
\includegraphics[width=0.25\textwidth,angle=-90]{f8b.eps}
\includegraphics[width=0.25\textwidth,angle=-90]{f8c.eps}
}
\center\mbox{
\includegraphics[width=0.25\textwidth,angle=-90]{f8d.eps}
\includegraphics[width=0.25\textwidth,angle=-90]{f8e.eps}
\includegraphics[width=0.25\textwidth,angle=-90]{f8f.eps}
}
\center\mbox{
\includegraphics[width=0.25\textwidth,angle=-90]{f8g.eps}
\includegraphics[width=0.25\textwidth,angle=-90]{f8h.eps}
\includegraphics[width=0.25\textwidth,angle=-90]{f8i.eps}
}
\caption[]{Images of GJ 4247 and its calibrators J2203+2811 (secondary) and J2205+2926
(primary).  Contours and labels are as described for 
Figure~\ref{fig:im65a}.\label{fig:im4247}}
\end{figure}

\begin{deluxetable}{lrrrrrrrlrrr}
\rotate
\tablewidth \textheight
\tabletypesize{\scriptsize}
\tablecaption{Stellar Data \label{tab:stardata}}

\tablehead{
\colhead{GJ}  & 
\colhead{RA} & 
\colhead{Dec} & 
\colhead{Pos. Err} &
\colhead{PM$_\alpha$,PM$_\delta$} & 
\colhead{PM$_{err}$} & 
\colhead{$\Pi$} & 
\colhead{$\Pi_{err}$} & 
\colhead{Sp} & 
\colhead{m$_B$} &  
\colhead{m$_V$} &  
\colhead{Log L$_X$} 
\\
 &
\colhead{(J2000)}& 
\colhead{(J2000)}& 
\colhead{(mas,mas,deg)} & 
\colhead{(mas/yr)}& 
\colhead{(mas/yr,mas/yr,deg)} & 
\colhead{(mas)}&  
\colhead{(mas)} &  
& 
& 
& 
\colhead{(erg s$^-1$)}
}

\startdata

1005A           & 00 15 27.70 & -16 -7 -56.00 & \dots,\dots,\dots & \dots, \dots & \dots, \dots, \dots & 191.80 &   0.00 & M4       & \dots &  13.29 &  26.21 \\ 
15A             & 00 18 22.89 & 44 01 22.63 & 06.48,05.55,  73 & 2888.70,  410.21 &    0.75,    0.63,   79 & 280.20 &   1.05 & M1.5V    &   9.63 &   8.07 &  26.43 \\ 
15B             & 00 18 25.81 & 44 01 38.00 & \dots,\dots,\dots & 2876.00,  341.00 &    5.00,    5.00,   97 & 280.20 &   0.00 & M3.5     &  12.84 &  11.04 &  27.38 \\ 
2005B           & 00 24 44.21 & -27 -8 -24.40 & \dots,\dots,\dots & \dots, \dots & \dots, \dots, \dots & 135.30 &   0.00 & M8.5V    & \dots & \dots &  27.32 \\ 
2005C           & 00 24 44.21 & -27 -8 -24.40 & \dots,\dots,\dots & \dots, \dots & \dots, \dots, \dots & 135.30 &   0.00 & M9V      & \dots & \dots &  27.32 \\ 
48              & 01 02 32.23 & 71 40 47.34 & 11.19,08.66, 131 & 1745.50, -380.80 &    1.25,    0.99,  135 & 122.70 &   1.23 & M3.5V:e  &  11.42 &   9.96 &  26.80 \\ 
53B             & 01 08 16.40 & 54 55 13.00 & \dots,\dots,\dots & 3420.00, -1600.00 & \dots, \dots, \dots & 132.40 &   0.00 & sdM      & \dots &  11.10 &  26.82 \\ 
3076            & 01 11 25.42 & 15 26 21.90 & \dots,\dots,\dots &  185.00, -121.00 &    5.00,    5.00,   90 & \dots & \dots & M5       & \dots &  13.59 &  27.70 \\ 
54.1            & 01 12 30.64 & -16 -59 -56.28 & 50.02,21.33,  58 & 1209.90,  646.88 &    5.93,    2.70,   57 & 269.00 &   7.57 & M4.5     &  12.80 &  11.60 &  27.40 \\ 
65A             & 01 39 01.54 & -17 -57 -1.80 & \dots,\dots,\dots & 3296.00,  563.00 &    5.00,    5.00,  142 & 373.70 &   0.00 & M5.5V:e  &  14.42 &  12.57 &  27.60 \\ 
65B             & 01 39 01.54 & -17 -57 -1.80 & \dots,\dots,\dots & 3296.00,  563.00 &    5.00,    5.00,  142 & 381.00 &   6.00 & M5.5e    &  14.37 &  12.52 &  27.60 \\ 
83.1            & 02 00 12.96 & 13 03 06.70 & \dots,\dots,\dots & 1091.00, -1780.00 &    5.00,    5.00,   45 & 222.00 &   5.00 & M4.5     &  14.08 &  12.26 &  27.38 \\ 
3125            & 02 01 54.08 & 73 32 32.00 & \dots,\dots,\dots &  276.00, -112.00 &    5.00,    5.00,   90 & \dots & \dots & M4.5     &  16.02 &  14.12 &  27.47 \\ 
84              & 02 05 04.85 & -17 -36 -52.67 & 16.44,10.73,  63 & 1317.50, -173.90 &    1.86,    1.21,   62 & 105.90 &   2.04 & M2.5     &  11.71 &  10.19 &  27.69 \\ 
3146            & 02 16 29.85 & 13 35 12.70 & \dots,\dots,\dots &  501.00, -437.00 &    5.00,    5.00,   58 & 118.20 &   6.80 & M5.5     &  17.77 &  15.79 &  27.26 \\ 
102             & 02 33 37.18 & 24 55 37.60 & \dots,\dots,\dots &   51.00, -679.00 &    5.00,    5.00,   77 & 129.00 &  12.00 & M4       &  14.66 &  12.96 &  28.00 \\ 
105C            & 02 35 58.80 & 06 52 01.00 & 3000.00,3000.00,  45 & \dots, \dots & \dots, \dots, \dots & 138.70 &   0.00 & M6V      & \dots & \dots &  27.34 \\ 
109             & 02 44 15.51 & 25 31 24.08 & 21.56,14.24, 100 &  864.67, -367.10 &    2.39,    1.66,   99 & 132.40 &   2.48 & M3       &  12.13 &  10.58 &  27.33 \\ 
3193B           & 03 01 51.42 & -16 -35 -36.13 & \dots,\dots,\dots & -343.00, -291.00 &    2.00,    2.00,   90 & \dots & \dots & M3       & \dots &  11.30 &  27.88 \\ 
144.0           & 03 32 55.84 & -9 -27 -29.74 & 09.48,06.79,  53 & -976.30,   17.98 &    1.09,    0.77,   49 & 310.70 &   0.85 & K2V      &   4.61 &   3.73 & \dots \\ 
166C            & 04 15 21.50 & -7 -39 -22.30 & \dots,\dots,\dots & -2239.00, -3419.00 & \dots, \dots, \dots & 198.20 &   0.00 & M4.5     &  12.85 &  11.17 &  28.16 \\ 
169.1A           & 04 31 11.52 & 58 58 37.46 & \dots,\dots,\dots & 1300.00, -2049.00 &    4.00,    3.00,    0 & 178.00 &   2.00 & M4       &  12.73 &  11.08 &  26.41 \\ 
3304            & 04 38 12.61 & 28 13 00.40 & \dots,\dots,\dots &  385.00,  -80.00 &    5.00,    5.00,   27 & \dots & \dots & M4       &  14.16 &  12.51 &  28.34 \\ 
176             & 04 42 55.78 & 18 57 29.42 & 29.29,13.32,  62 &  659.78, -1114.00 &    3.45,    1.48,   61 & 106.10 &   2.51 & K5       &  11.48 &   9.97 & \dots \\ 
3323            & 05 01 57.43 & -6 -56 -46.50 & \dots,\dots,\dots & -550.00, -533.00 &    5.00,    5.00,   90 & 163.00 &   0.00 & M4       &  13.92 &  12.16 &  27.48 \\ 
185B            & 05 02 28.42 & -21 -15 -23.90 & \dots,\dots,\dots & \dots, \dots & \dots, \dots, \dots & 129.40 &   0.00 & M2V      & \dots & \dots &  27.12 \\ 
190             & 05 08 35.05 & -18 -10 -19.37 & 13.47,09.72, 153 &  503.76, -1400.00 &    1.52,    1.15,  152 & 107.30 &   2.00 & M3.5     &  11.83 &  10.31 &  27.28 \\ 
205             & 05 31 27.40 & -3 -40 -38.02 & 10.23,06.46,  81 &  763.03, -2092.00 &    1.16,    0.72,   79 & 175.70 &   1.20 & M1.5V    &   9.44 &   7.92 &  27.66 \\ 
213             & 05 42 09.27 & 12 29 21.62 & 22.77,13.49,  89 & 1998.30, -1570.00 &    2.67,    1.67,   88 & 172.70 &   3.88 & M4V      &  13.15 &  11.48 &  27.64 \\ 
3379            & 06 00 03.49 & 02 42 23.67 & \dots,\dots,\dots &  311.00,  -42.00 &    2.00,    2.00,    0 & 186.30 &   6.20 & M3.5V    &  13.01 &  11.33 &  27.92 \\ 
226             & 06 10 19.85 & 82 06 24.31 & 11.60,09.65,  47 &   50.24, -1336.00 &    1.34,    1.11,   46 & 106.40 &   1.44 & M2       &  12.00 &  10.49 &  27.09 \\ 
229A            & 06 10 34.62 & -21 -51 -52.72 & 08.24,04.82, 156 & -137.00, -714.00 &    0.93,    0.54,  158 & 173.10 &   1.12 & M1/M2V   &   9.65 &   8.14 &  27.11 \\ 
234A            & 06 29 23.40 & -2 -48 -50.32 & 28.81,15.90,  55 &  694.66, -618.50 &    3.43,    1.84,   53 & 242.80 &   2.64 & M4.5     &  13.08 &  11.10 &  27.97 \\ 
234B            & 06 29 23.52 & -2 -48 -51.10 & 2000.00,2000.00,  45 &  707.00, -703.00 & \dots, \dots, \dots & 242.80 &   0.00 & M8V      &  16.00 &  14.60 &  27.97 \\ 
250B            & 06 52 18.07 & -5 -11 -25.60 & 2000.00,2000.00,  89 & -541.00,    0.00 & \dots, \dots, \dots & \dots & \dots & M2       &  11.47 &  10.05 &  28.10 \\ 
251             & 06 54 48.96 & 33 16 05.43 & 18.67,10.28,  72 & -729.20, -399.20 &    2.13,    1.21,   70 & 181.30 &   1.87 & M3       &  11.47 &   9.89 &  26.11 \\ 
268A            & 07 10 01.83 & 38 31 46.06 & 46.84,22.32,  80 & -439.60, -948.20 &    5.40,    2.63,   78 & 157.20 &   3.32 & M4.5     &  13.19 &  11.47 &  27.87 \\ 
273             & 07 27 24.50 & 05 13 32.83 & 12.10,07.20, 103 &  571.26, -3694.00 &    1.43,    0.87,  102 & 263.20 &   1.43 & M3.5     &  11.42 &   9.89 &  26.55 \\ 
285             & 07 44 40.17 & 03 33 08.83 & 22.44,14.81,  98 & -344.80, -450.80 &    2.62,    1.73,   96 & 168.50 &   2.67 & M4.5V:e  &  12.73 &  11.12 &  28.57 \\ 
300             & 08 12 40.87 & -21 -33 -6.80 & \dots,\dots,\dots &    9.00, -694.00 &    5.00,    5.00,   90 & 166.00 &  11.00 & M4       &  13.80 &  13.00 &  26.26 \\ 
2066            & 08 16 07.98 & 01 18 09.26 & 18.91,13.34,  89 & -375.00,   60.11 &    2.17,    1.49,   88 & 109.20 &   1.79 & M2       &  11.55 &  10.05 & \dots \\ 
1111            & 08 29 49.35 & 26 46 33.70 & \dots,\dots,\dots & -1111.00, -612.00 &    5.00,    5.00,  152 & 275.80 &   0.00 & M6       &  16.87 &  14.81 &  26.58 \\ 
3517            & 08 53 36.20 & -3 -29 -32.10 & 60.00,60.00,  45 & -502.00, -211.00 &    5.00,    5.00,    4 & 116.80 &   1.50 & M9V      &  20.80 &  20.80 &  26.67 \\ 
1116A           & 08 58 15.19 & 19 45 47.10 & 70.00,60.00,   0 & -846.00,  -65.00 &    5.00,    5.00,    0 & 190.00 &   3.00 & M8Ve     &  15.90 &  14.06 &  27.82 \\ 
1116B           & 08 58 15.13 & 19 45 47.10 & \dots,\dots,\dots & -846.00,  -65.00 &    5.00,    5.00,    0 & 190.00 &   5.00 & M5.5     &  16.85 &  14.92 &  27.82 \\ 
3522            & 08 58 56.45 & 08 28 24.60 & 2000.00,2000.00,  97 &  410.00, -367.00 &  100.00,  100.00,   90 & \dots & \dots & M3.5     &  12.57 &  10.89 &  28.65 \\ 
338A            & 09 14 22.79 & 52 41 11.85 & 57.93,32.48,  91 & -1533.00, -562.90 &    6.80,    3.97,   91 & 161.50 &   5.23 & M0V      &   9.07 &   7.64 & \dots \\ 
382             & 10 12 17.67 & -3 -44 -44.38 & 11.74,07.96, 118 & -152.90, -242.80 &    1.38,    0.93,  116 & 127.90 &   1.53 & M1.5     &  10.77 &   9.26 &  27.45 \\ 
388             & 10 19 36.27 & 19 52 11.90 & 137.63,133.11,  90 & -500.80,  -46.00 &    2.00,    1.90,   88 & 213.00 &   4.00 & M3.5V    &  10.97 &   9.43 &  28.80 \\ 
393             & 10 28 55.55 & 00 50 27.60 & 20.08,11.93, 136 & -602.30, -731.80 &    2.49,    1.37,  139 & 138.20 &   2.13 & M2       &  11.15 &   9.63 & \dots \\ 
3622            & 10 48 12.61 & -11 -20 -9.70 & \dots,\dots,\dots &  580.00, -1531.00 &    5.00,    5.00,  122 & 221.00 &   3.60 & M6.5     &  17.70 &  15.60 &  25.96 \\ 
406             & 10 56 28.99 & 07 00 52.00 & 2000.00,2000.00,   0 & -3842.00, -2725.00 &  100.00,  100.00,    1 & 425.00 &   7.00 & M5.5V    &  15.54 &  13.54 &  26.97 \\ 
408             & 11 00 04.26 & 22 49 58.67 & 14.22,08.99,  90 & -426.20, -279.90 &    1.69,    1.04,   85 & 150.90 &   1.59 & M2.5     &  11.58 &  10.03 & \dots \\ 
411             & 11 03 20.19 & 35 58 11.55 & 08.43,04.71, 134 & -580.40, -4769.00 &    0.95,    0.53,  133 & 392.50 &   0.91 & M2V      &   9.00 &   7.49 &  26.89 \\ 
412A            & 11 05 28.58 & 43 31 36.39 & 08.89,06.78, 123 & -4409.00,  942.31 &    1.02,    0.77,  127 & 206.90 &   1.19 & M0.5     &  10.22 &   8.68 &  27.49 \\ 
412B            & 11 05 30.90 & 43 31 17.90 & \dots,\dots,\dots & -4328.00,  952.00 &    5.00,    5.00,   90 & 206.90 &   0.00 & M6       &  16.45 &  14.45 &  27.49 \\ 
424             & 11 20 04.83 & 65 50 47.35 & 08.82,07.38, 111 & -2946.00,  183.68 &    1.00,    0.86,  112 & 109.90 &   1.11 & M0       &  10.74 &   9.32 &  27.04 \\ 
436             & 11 42 11.09 & 26 42 23.65 & 25.12,14.95,  80 &  896.34, -813.70 &    2.96,    1.79,   80 &  97.73 &   2.27 & M2.5     &  12.20 &  10.68 & \dots \\ 
445             & 11 47 41.38 & 78 41 28.18 & 13.54,10.69,  81 &  743.58,  480.47 &    1.56,    1.20,   81 & 185.50 &   1.43 & M3.5     &  12.37 &  10.78 &  26.65 \\ 
450             & 11 51 07.34 & 35 16 19.26 & 10.81,06.77,  87 & -271.90,  254.93 &    1.30,    0.83,   88 & 116.90 &   1.38 & M1       &  11.30 &   9.78 &  27.56 \\ 
451B            & 11 52 60.00 & 37 43 00.00 & 180004.00,180003.00,  55 & 3995.00, -5806.00 &   25.00,   25.00,  145 & \dots & \dots & M5.5V    & \dots &  12.00 &  26.49 \\ 
1156            & 12 18 59.49 & 11 07 32.80 & 2000.00,2000.00,  90 & -1246.00,  186.00 &  100.00,  100.00,   90 & 152.90 &   0.00 & M5       &  15.62 &  13.79 &  27.60 \\ 
473A            & 12 33 16.30 & 09 01 26.00 & \dots,\dots,\dots & \dots, \dots & \dots, \dots, \dots & 227.90 &   0.00 & M4V      & \dots &  12.44 &  27.52 \\ 
473B            & 12 33 19.10 & 09 01 10.00 & 3000.00,3000.00, 179 & -1744.00,  263.00 &   25.00,   25.00,   58 & 227.90 &   0.00 & M7       & \dots &  13.40 &  27.52 \\ 
486             & 12 47 56.62 & 09 45 05.03 & 26.79,13.07, 111 & -1007.00, -461.00 &    3.03,    1.57,  113 & 121.70 &   2.90 & M3.5     &  12.96 &  11.40 & \dots \\ 
493.1           & 13 00 33.54 & 05 41 08.50 & 2000.00,2000.00,  45 & -940.00,  241.00 &  100.00,  100.00,  178 & \dots & \dots & M4.5     &  15.12 &  13.37 &  27.82 \\ 
504             & 13 16 46.52 & 09 25 26.96 & 07.43,04.09, 111 & -334.50,  190.70 &    0.85,    0.46,  113 &  55.71 &   0.85 & G0Vs     &   5.81 &   5.22 & \dots \\ 
514             & 13 29 59.79 & 10 22 37.78 & 10.61,06.10,  91 & 1127.90, -1074.00 &    1.17,    0.67,   94 & 131.10 &   1.29 & M0.5     &  10.54 &   9.04 &  27.38 \\ 
3789            & 13 31 46.61 & 29 16 36.69 & \dots,\dots,\dots & -234.00, -137.00 &    2.00,    2.00,   90 & 126.00 &   0.00 & M4       &  13.52 &  11.95 &  28.23 \\ 
526             & 13 45 43.78 & 14 53 29.47 & 09.87,06.83, 128 & 1778.30, -1455.00 &    1.09,    0.75,  127 & 184.10 &   1.27 & M2V      &   9.91 &   8.46 &  26.87 \\ 
3820            & 13 59 10.43 & -19 -50 -3.60 & \dots,\dots,\dots & -552.00, -183.00 &    5.00,    5.00,   11 & \dots & \dots & M4       &  14.70 &  13.00 &  28.04 \\ 
557.0           & 14 34 40.82 & 29 44 42.47 & 06.42,04.08, 140 &  188.32,  132.72 &    0.73,    0.47,  137 &  64.66 &   0.72 & F2V      &   4.82 &   4.46 & \dots \\ 
566A            & 14 51 23.10 & 19 06 02.00 & 3250.00,3250.00,   0 &  139.00,  -99.00 &   25.00,   25.00,  157 & 149.20 &   0.00 & G8V      &   5.40 &   4.70 & \dots \\ 
569B            & 14 54 29.00 & 16 06 05.00 & \dots,\dots,\dots & \dots, \dots & \dots, \dots, \dots & \dots & \dots & M8.5     & \dots &  18.10 &  28.53 \\ 
569A            & 14 54 29.24 & 16 06 03.82 & 16.58,11.37,  84 &  275.95, -122.10 &    1.96,    1.31,   80 & 101.90 &   1.67 & M2.5V    &  11.68 &  10.20 &  28.53 \\ 
3877            & 14 56 38.31 & -28 -9 -47.40 & 60.00,60.00,  17 & -497.00, -827.00 &    5.00,    5.00,   22 & 157.80 &   5.10 & M5V      &  18.39 &  17.05 &  26.19 \\ 
570C            & 14 57 26.50 & -21 -24 -41.00 & \dots,\dots,\dots & \dots, \dots & \dots, \dots, \dots & 133.60 &  33.56 & M3V      & \dots &   9.94 &  27.74 \\ 
570B            & 14 57 26.54 & -21 -24 -41.47 & 404.97,242.74,  99 &  987.02, -1666.00 &   44.89,   28.17,   98 & 133.60 &  33.56 & M1V      &   9.68 &   8.10 &  27.74 \\ 
581             & 15 19 26.82 & -7 -43 -20.21 & 27.21,14.51,  77 & -1224.00,  -99.51 &    3.13,    1.68,   76 & 159.50 &   2.27 & M3       &  12.17 &  10.56 & \dots \\ 
625             & 16 25 24.62 & 54 18 14.77 & 11.85,10.46,  63 &  432.29, -170.70 &    1.34,    1.18,   67 & 151.90 &   1.11 & M1.5     &  10.89 &  10.40 &  26.77 \\ 
628             & 16 30 18.06 & -12 -39 -45.34 & 22.31,11.34, 124 &  -93.62, -1184.00 &    2.51,    1.25,  123 & 234.50 &   1.82 & M3.5     &  11.72 &  10.12 &  26.79 \\ 
638             & 16 45 06.35 & 33 30 33.23 & 07.87,06.41, 173 &  -39.18,  383.43 &    0.89,    0.72,  170 & 102.30 &   0.88 & K5       &   9.48 &   8.11 & \dots \\ 
643             & 16 55 25.23 & -8 -19 -21.27 & 39.55,23.03,  84 & -813.40, -895.10 &    4.43,    2.46,   82 & 153.90 &   4.04 & M3.5     &  13.40 &  11.70 &  29.08 \\ 
644A            & 16 55 26.34 & -8 -20 -53.90 & 1285.00,1285.00,   9 & -791.00, -878.00 &   25.00,   25.00,   90 & 153.90 &   0.00 & M3       &  11.40 &   9.80 &  29.08 \\ 
644B            & 16 55 26.34 & -8 -20 -53.90 & 1285.00,1285.00,   9 & -791.00, -878.00 &   25.00,   25.00,   90 & 153.90 &   0.00 & M4       & \dots &   9.80 &  29.08 \\ 
644C            & 16 55 35.29 & -8 -23 -40.10 & 60.00,60.00,  87 & -771.00, -871.00 &   23.00,   21.00,    0 & 153.90 &   0.00 & M6.5V    &  18.70 &  16.70 &  27.45 \\ 
1207            & 16 57 05.80 & -4 -20 -57.00 & 2000.00,2000.00,  45 &  513.00, -377.00 & \dots, \dots, \dots & \dots & \dots & M3.5     &  13.92 &  12.33 &  28.39 \\ 
661A            & 17 12 07.83 & 45 39 57.68 & \dots,\dots,\dots &  252.00, -1571.00 &    1.00,    1.00,   72 & 158.10 &   0.00 & M3.5     &  11.50 &  10.00 &  27.23 \\ 
661B            & 17 12 07.83 & 45 39 57.68 & \dots,\dots,\dots &  252.00, -1571.00 &    1.00,    1.00,   72 & 158.10 &   0.00 & M3       &  11.30 &  10.30 &  27.89 \\ 
673             & 17 25 45.23 & 02 06 41.12 & 07.17,05.06,  81 & -580.50, -1184.00 &    0.81,    0.57,   82 & 129.50 &   0.95 & K5       &   8.90 &   7.54 & \dots \\ 
687             & 17 36 25.90 & 68 20 20.91 & 11.63,08.60,  22 & -320.40, -1269.00 &    1.36,    1.02,   21 & 220.80 &   0.92 & M3       &  10.65 &   9.15 & \dots \\ 
686             & 17 37 53.35 & 18 35 30.15 & 12.63,09.72, 143 &  926.84,  983.18 &    1.42,    1.07,  142 & 123.00 &   1.62 & M1       &  11.15 &   9.62 & \dots \\ 
694             & 17 43 55.96 & 43 22 43.00 & 11.96,09.36,  48 &    9.48, -602.60 &    1.39,    1.08,   45 & 105.30 &   1.15 & M2.5     &  12.05 &  10.50 &  27.18 \\ 
695B            & 17 46 25.16 & 27 43 00.74 & 439.88,425.86,  90 & -332.00, -753.00 &    5.80,    5.60,   90 & \dots & \dots & M3.5     &  11.27 &   9.78 &  28.00 \\ 
695C            & 17 46 25.20 & 27 43 01.00 & \dots,\dots,\dots & -301.00, -750.00 &   25.00,   25.00,    0 & \dots & \dots & M4       & \dots &  10.80 &  28.00 \\ 
699             & 17 57 48.50 & 04 41 36.25 & 14.24,10.45,  67 & -798.70, 10337.00 &    1.66,    1.22,   67 & 549.30 &   1.58 & M4Ve     &  11.28 &   9.54 &  25.85 \\ 
701             & 18 05 07.58 & -3 -1 -52.75 & 10.60,08.53,  90 &  570.10, -332.50 &    1.22,    1.00,   90 & 128.20 &   1.44 & M1       &  10.89 &   9.37 &  27.13 \\ 
1224            & 18 07 32.85 & -15 -57 -47.10 & \dots,\dots,\dots & -617.00, -342.00 &    5.00,    5.00,  176 & 132.60 &   0.00 & M4.5     &  15.44 &  13.64 &  27.93 \\ 
4053            & 18 18 57.40 & 66 11 32.40 & 2000.00,2000.00,  45 &  464.00, -469.00 &  100.00,  100.00,   45 & 137.30 &   5.30 & M4.5     &  15.29 &  13.46 &  27.33 \\ 
4063            & 18 34 36.65 & 40 07 26.44 & \dots,\dots,\dots &   57.00, -203.00 &    2.00,    2.00,   90 & 138.00 &  40.00 & M3.5     &  12.84 &  11.42 &  26.92 \\ 
1230A           & 18 41 09.86 & 24 47 13.90 & 2000.00,2000.00,   2 &  517.00,   55.00 &  100.00,  100.00,   45 & \dots & \dots & M4.5     &  14.00 &  12.40 &  27.94 \\ 
1230B           & 18 41 09.86 & 24 47 13.90 & 2000.00,2000.00,   2 &  517.00,   53.00 &  100.00,  100.00,   45 & \dots & \dots & M5       &  16.10 &  14.40 &  27.94 \\ 
725A            & 18 42 46.69 & 59 37 49.42 & 32.44,27.69, 168 & -1327.00, 1801.90 &    3.61,    3.06,  164 & 280.20 &   2.57 & M3V      &  10.45 &   8.91 &  26.65 \\ 
725B            & 18 42 46.90 & 59 37 36.65 & 110.99,102.93, 162 & -1393.00, 1845.50 &   12.19,   11.32,  153 & 284.40 &   5.01 & M3.5     &  11.28 &   9.69 &  26.65 \\ 
729             & 18 49 49.36 & -23 -50 -10.44 & 19.73,12.25,  94 &  637.57, -192.40 &    2.23,    1.44,   94 & 336.40 &   1.82 & M3.5     &  12.55 &  10.95 &  27.78 \\ 
747A            & 19 07 43.00 & 32 32 41.30 & \dots,\dots,\dots & 1235.00, 1130.00 &    5.00,    5.00,  125 & 123.00 &   4.00 & M3       &  13.30 &  11.70 &  27.08 \\ 
747B            & 19 07 45.00 & 32 32 54.00 & 18043.00,18043.00, 138 & 1213.00, 1095.00 &   25.00,   25.00,   48 & \dots & \dots & M5       & \dots &  12.10 &  27.08 \\ 
752A            & 19 16 55.26 & 05 10 08.05 & 11.77,07.86,  87 & -578.80, -1331.00 &    1.33,    0.91,   89 & 170.20 &   1.37 & M2.5     &  10.63 &   9.13 &  26.94 \\ 
752B            & 19 16 57.66 & 05 09 00.40 & 2000.00,2000.00, 179 & -592.00, -1400.00 &  100.00,  100.00,   45 & 170.20 &   0.00 & M8V      &  19.42 &  17.30 &  27.05 \\ 
1245A           & 19 53 54.48 & 44 24 53.30 & \dots,\dots,\dots &  443.00, -581.00 &   20.00,   20.00,  112 & 215.00 &   3.00 & M5.5     &  15.31 &  13.41 &  27.47 \\ 
1245C           & 19 53 54.90 & 44 24 54.00 & 3000.00,3000.00,  45 & \dots, \dots & \dots, \dots, \dots & 220.20 &   0.00 & M5.5     & \dots &  17.50 &  27.47 \\ 
1245B           & 19 53 55.14 & 44 24 54.10 & \dots,\dots,\dots &  443.00, -581.00 &   20.00,   20.00,  112 & 206.00 &   3.00 & M5.5     &  15.97 &  13.99 &  27.47 \\ 
791.2           & 20 29 48.41 & 09 41 19.30 & 2000.00,2000.00,   0 &  704.00,   92.00 &  100.00,  100.00,   94 & 112.00 &   3.00 & M4.5V    &  14.69 &  13.04 &  27.89 \\ 
793             & 20 30 32.05 & 65 26 58.42 & 09.95,07.43,  29 &  443.24,  284.03 &    1.15,    0.87,   29 & 125.60 &   1.11 & M2.5     &  12.00 &  10.44 &  27.77 \\ 
803             & 20 45 09.53 & -31 -20 -27.24 & 14.87,06.92,  65 &  280.37, -360.00 &    1.68,    0.79,   65 & 100.50 &   1.35 & M1Ve     &  10.05 &   8.61 & \dots \\ 
809             & 20 53 19.79 & 62 09 15.81 & 07.04,05.18,  30 &    1.08, -774.20 &    0.80,    0.62,   28 & 141.90 &   0.77 & M0.5     &  10.03 &   8.54 &  27.47 \\ 
829             & 21 29 36.81 & 17 38 35.85 & 24.71,09.75,  86 & 1008.00,  376.20 &    2.91,    1.12,   85 & 148.20 &   1.85 & M3.5     &  11.96 &  10.35 & \dots \\ 
831A            & 21 31 17.80 & -9 -47 -25.00 & \dots,\dots,\dots & \dots, \dots & \dots, \dots, \dots & \dots & \dots & M4.5     & \dots &  12.60 &  27.45 \\ 
4247            & 22 01 13.12 & 28 18 24.86 & \dots,\dots,\dots &  372.11,   36.48 &    3.51,    2.59,   23 & 111.50 &   3.19 & M4       &  13.90 &  11.99 &  28.36 \\ 
849             & 22 09 40.35 & -4 -38 -26.62 & 24.38,11.58,  93 & 1134.90,  -19.71 &    2.75,    1.30,   91 & 113.90 &   2.10 & M3.5     &  11.94 &  10.42 & \dots \\ 
4274            & 22 23 06.99 & -17 -36 -26.40 & \dots,\dots,\dots &  248.00, -895.00 &  231.00,   64.00,  180 & 134.10 &   5.60 & M4.5     &  15.09 &  13.25 &  27.80 \\ 
860A            & 22 27 59.47 & 57 41 45.15 & 27.09,25.47,  64 & -870.30, -471.20 &    3.06,    2.91,   48 & 249.50 &   3.03 & M3       &  11.24 &   9.59 &  27.72 \\ 
860B            & 22 28 00.34 & 57 41 44.40 & 2000.00,2000.00,  85 & -711.00, -320.00 &  100.00,  100.00,   80 & 249.50 &   0.00 & M4       &  12.90 &  10.30 &  27.72 \\ 
866             & 22 38 33.62 & -15 -17 -59.20 & \dots,\dots,\dots & 2314.00, 2295.00 &    5.00,    5.00,   13 & 300.00 &   5.00 & M5.5     &  14.14 &  12.18 &  27.23 \\ 
867B            & 22 38 45.31 & -20 -36 -52.10 & \dots,\dots,\dots &  427.00,  -61.00 &    7.00,    7.00,   68 & \dots & \dots & M3.5     &  13.04 &  11.43 &  29.31 \\ 
867A            & 22 38 45.58 & -20 -37 -16.08 & 108000.00,108000.00, 255 &  450.59,  -79.86 & \dots, \dots, \dots & 115.70 &   1.50 & M1.5     &  10.57 &   9.07 &  29.31 \\ 
873             & 22 46 49.73 & 44 20 02.37 & 11.83,09.48, 149 & -704.60, -459.40 &    1.40,    1.10,  148 & 198.00 &   2.05 & M3.5e    &  11.45 &  10.09 &  28.99 \\ 
876A            & 22 53 16.73 & -14 -15 -49.32 & 32.97,13.91,  95 &  960.31, -675.60 &    3.77,    1.58,   97 & 212.60 &   2.10 & M4       &  11.77 &  10.17 &  26.49 \\ 
880             & 22 56 34.81 & 16 33 12.36 & 13.63,07.01,  31 & -1033.00, -283.30 &    1.60,    0.80,   30 & 145.20 &   1.22 & M1.5V    &  10.17 &   8.66 &  27.09 \\ 
896A            & 23 31 52.18 & 19 56 14.13 & 19.96,11.30,  60 &  554.40,  -62.61 &    2.25,    1.28,   56 & 160.00 &   2.81 & M3.5     &  11.51 &  10.32 &  29.06 \\ 
896B            & 23 31 52.56 & 19 56 13.90 & \dots,\dots,\dots &  602.00,   17.00 &   14.00,    9.00,    0 & 160.00 &   0.00 & M4.5     &  14.40 &  12.40 &  29.06 \\ 
905             & 23 41 55.01 & 44 10 38.90 & \dots,\dots,\dots &  100.00, -1594.00 &    5.00,    5.00,  173 & 315.00 &   2.00 & M5.5V    &  14.19 &  12.28 &  27.10 \\ 
1289            & 23 43 06.28 & 36 32 14.00 & 2000.00,2000.00,  45 &  930.00, -136.00 &  100.00,  100.00,    0 & \dots & \dots & M4       &  14.27 &  12.67 &  27.69 \\ 
4360            & 23 45 31.27 & -16 -10 -19.30 & 2000.00,2000.00, 177 & -395.00, -558.00 & \dots, \dots, \dots & \dots & \dots & M5       &  14.80 &  14.50 &  27.66 \\ 
908             & 23 49 12.53 & 02 24 04.40 & 19.28,06.30,  81 &  995.31, -968.40 &    2.33,    0.77,   82 & 167.50 &   1.49 & M1       &  10.46 &   8.98 &  27.12 \\ 
\enddata
\end{deluxetable}

\begin{deluxetable}{lrr}
\tablecaption{Detected Stars \label{tab:detect}}
\tablehead{
\colhead{GJ} &
\colhead{Epoch} &
\colhead{Flux Density} \\
&
\colhead{(YYYYMMDD)} &
\colhead{($\mu$Jy)}
}
\startdata

 53B & 20050825 & $ 350 \pm   86$ \\ 
 65B & 20050807 & $4271 \pm   66$ \\ 
 84 & 20050807 & $ 196 \pm   57$ \\ 
 102 & 20050807 & $ 182 \pm   49$ \\ 
 \dots & 20050825 & $< 204$ \\ 
 109 & 20050612 & $ 131 \pm   40$ \\ 
 \dots & 20050709 & $< 138$ \\ 
 412B & 20050612 & $ 152 \pm   45$ \\ 
 \dots & 20050711 & $< 189$ \\ 
 \dots & 20050903 & $1300 \pm   66$ \\ 
 557 & 20050612 & $ 236 \pm   68$ \\ 
 644C & 20050816 & $< 171$ \\ 
 \dots & 20050824 & $ 224 \pm   57$ \\ 
 661AB & 20050816 & $ 536 \pm   52$ \\ 
 \dots & 20050822 & $< 147$ \\ 
 686 & 20050609 & $ 122 \pm   39$ \\ 
 \dots & 20050711 & $< 147$ \\ 
 729 & 20050816 & $ 261 \pm   59$ \\ 
 \dots & 20050822 & $ 220 \pm   52$ \\ 
 747A & 20050809 & $ 450 \pm  133$ \\ 
 \dots & 20050824 & $< 255$ \\ 
 803 & 20050609 & $1232 \pm   50$ \\ 
 \dots & 20050612 & $ 368 \pm   53$ \\ 
 \dots & 20050711 & $ 222 \pm   59$ \\ 
 \dots & 20050729 & $ 324 \pm   49$ \\ 
 866 & 20050709 & $ 158 \pm   46$ \\ 
 867B & 20050824 & $ 376 \pm   64$ \\ 
 873 & 20050824 & $ 546 \pm   90$ \\ 
 896A & 20050807 & $ 567 \pm   52$ \\ 
 \dots & 20050825 & $1027 \pm   62$ \\ 
 1005A & 20050807 & $< 204$ \\ 
 \dots & 20050825 & $ 348 \pm   96$ \\ 
 1116AB & 20050903 & $1482 \pm   82$ \\ 
 1207 & 20050816 & $ 780 \pm   63$ \\ 
 \dots & 20050824 & $ 195 \pm   58$ \\ 
 1224 & 20050816 & $1271 \pm   69$ \\ 
 \dots & 20050824 & $ 185 \pm   59$ \\ 
 1230AB & 20050809 & $< 171$ \\ 
 \dots & 20050824 & $ 199 \pm   61$ \\ 
 2066 & 20050709 & $ 364 \pm   55$ \\ 
 \dots & 20050711 & $< 177$ \\ 
 3146 & 20050809 & $< 267$ \\ 
 \dots & 20050825 & $ 281 \pm   71$ \\ 
 3789 & 20050816 & $6203 \pm  105$ \\ 
 \dots & 20050824 & $3139 \pm   65$ \\ 
 4063 & 20050822 & $< 150$ \\ 
 \dots & 20050816 & $ 219 \pm   53$ \\ 
 4247 & 20050816 & $3054 \pm   75$ \\ 
 \dots & 20050822 & $ 791 \pm   66$ \\ 
 4360 & 20050807 & $2174 \pm  130$ \\ 
 \dots & 20050825 & $< 300$ \\ 
\enddata
\end{deluxetable}

\begin{deluxetable}{lrr}
\tablecaption{Undetected Stars \label{tab:nodetect}}
\tablehead{
\colhead{GJ} &
\colhead{Epoch} &
\colhead{Flux Density Limit} \\
&
\colhead{(YYYYMMDD)} &
\colhead{($\mu$Jy)}
}
\startdata
 15AB & 20050612 & $< 123$ \\ 
 \dots & 20050709 & $< 144$ \\ 
 48 & 20050612 & $< 132$ \\ 
 \dots & 20050709 & $< 147$ \\ 
 54.1 & 20050807 & $< 567$ \\ 
 83.1 & 20050809 & $< 243$ \\ 
 \dots & 20050825 & $< 210$ \\ 
 105C & 20050807 & $< 201$ \\ 
 \dots & 20050825 & $< 231$ \\ 
 144.0 & 20050709 & $< 144$ \\ 
 \dots & 20050711 & $< 105$ \\ 
 166C & 20050807 & $< 618$ \\ 
 169.1A & 20050809 & $< 234$ \\ 
 \dots & 20050825 & $< 231$ \\ 
 176 & 20050709 & $< 216$ \\ 
 \dots & 20050711 & $< 255$ \\ 
 205 & 20050709 & $< 153$ \\ 
 \dots & 20050711 & $< 144$ \\ 
 226 & 20050612 & $< 141$ \\ 
 \dots & 20050709 & $< 156$ \\ 
 251 & 20050722 & $< 180$ \\ 
 273 & 20050709 & $< 156$ \\ 
 \dots & 20050711 & $< 162$ \\ 
 338A & 20050709 & $< 156$ \\ 
 \dots & 20050711 & $< 120$ \\ 
 382 & 20050722 & $< 147$ \\ 
 388 & 20050903 & $< 567$ \\ 
 393 & 20050722 & $< 156$ \\ 
 406 & 20050903 & $< 735$ \\ 
 408 & 20050612 & $< 126$ \\ 
 \dots & 20050722 & $< 141$ \\ 
 411 & 20050612 & $< 138$ \\ 
 \dots & 20050722 & $< 186$ \\ 
 \dots & 20050903 & $< 192$ \\ 
 412A & 20050612 & $< 141$ \\ 
 \dots & 20050711 & $< 189$ \\ 
 \dots & 20050903 & $< 198$ \\ 
 424 & 20050612 & $< 225$ \\ 
 \dots & 20050709 & $< 180$ \\ 
 \dots & 20050711 & $< 192$ \\ 
 \dots & 20050903 & $< 204$ \\ 
 436 & 20050612 & $< 138$ \\ 
 \dots & 20050722 & $< 156$ \\ 
 445 & 20050612 & $< 171$ \\ 
 \dots & 20050903 & $< 225$ \\ 
 450 & 20050612 & $< 132$ \\ 
 \dots & 20050903 & $< 189$ \\ 
 451B & 20050816 & $< 393$ \\ 
 \dots & 20050824 & $< 243$ \\ 
 473AB & 20050824 & $< 180$ \\ 
 486 & 20050612 & $< 129$ \\ 
 493.1 & 20050824 & $< 183$ \\ 
 504 & 20050612 & $< 159$ \\ 
 \dots & 20050722 & $< 144$ \\ 
 514 & 20050612 & $< 117$ \\ 
 \dots & 20050722 & $< 132$ \\ 
 526 & 20050612 & $< 177$ \\ 
 \dots & 20050722 & $< 153$ \\ 
 566A & 20050609 & $< 141$ \\ 
 \dots & 20050722 & $< 144$ \\ 
 569AB & 20050609 & $< 147$ \\ 
 570B & 20050824 & $< 285$ \\ 
 \dots & 20050824 & $< 285$ \\ 
 581 & 20050612 & $< 117$ \\ 
 625 & 20050612 & $< 129$ \\ 
 \dots & 20050711 & $< 153$ \\ 
 628 & 20050816 & $< 252$ \\ 
 \dots & 20050824 & $< 192$ \\ 
 638 & 20050612 & $< 120$ \\ 
 \dots & 20050711 & $< 153$ \\ 
 643 & 20050816 & $< 189$ \\ 
 \dots & 20050824 & $< 171$ \\ 
 644AB & 20050816 & $< 171$ \\ 
 \dots & 20050824 & $< 171$ \\ 
 673 & 20050609 & $< 111$ \\ 
 \dots & 20050711 & $< 156$ \\ 
 687 & 20050612 & $< 126$ \\ 
 \dots & 20050709 & $< 162$ \\ 
 694 & 20050612 & $< 129$ \\ 
 \dots & 20050711 & $< 150$ \\ 
 695BC & 20050816 & $< 270$ \\ 
 \dots & 20050822 & $< 240$ \\ 
 699 & 20050609 & $< 123$ \\ 
 \dots & 20050711 & $< 153$ \\ 
 701 & 20050609 & $< 123$ \\ 
 \dots & 20050711 & $< 171$ \\ 
 725A & 20050612 & $< 141$ \\ 
 \dots & 20050709 & $< 168$ \\ 
 725B & 20050612 & $< 141$ \\ 
 \dots & 20050709 & $< 168$ \\ 
 747B & 20050809 & $< 441$ \\ 
 \dots & 20050824 & $< 255$ \\ 
 752AB & 20050609 & $< 144$ \\ 
 \dots & 20050711 & $< 171$ \\ 
 791.2 & 20050822 & $< 210$ \\ 
 793 & 20050612 & $< 123$ \\ 
 \dots & 20050709 & $< 159$ \\ 
 809 & 20050612 & $< 153$ \\ 
 \dots & 20050709 & $< 189$ \\ 
 829 & 20050609 & $< 144$ \\ 
 \dots & 20050711 & $< 267$ \\ 
 831A & 20050825 & $< 231$ \\ 
 849 & 20050609 & $<  27$ \\ 
 \dots & 20050612 & $< 273$ \\ 
 860AB & 20050612 & $< 216$ \\ 
 \dots & 20050709 & $< 225$ \\ 
 867A & 20050824 & $< 192$ \\ 
 876A & 20050609 & $< 219$ \\ 
 \dots & 20050612 & $< 201$ \\ 
 \dots & 20050709 & $< 186$ \\ 
 880 & 20050609 & $< 129$ \\ 
 \dots & 20050612 & $< 114$ \\ 
 \dots & 20050709 & $< 162$ \\ 
 896B & 20050807 & $< 162$ \\ 
 \dots & 20050825 & $< 186$ \\ 
 905 & 20050824 & $< 174$ \\ 
 908 & 20050612 & $< 105$ \\ 
 \dots & 20050709 & $< 123$ \\ 
 1156 & 20050824 & $< 243$ \\ 
 1245ABC & 20050809 & $< 219$ \\ 
 \dots & 20050825 & $< 192$ \\ 
 1289 & 20050824 & $< 459$ \\ 
 2005BC & 20050807 & $<1254$ \\ 
 \dots & 20050825 & $< 420$ \\ 
 3076 & 20050807 & $< 537$ \\ 
 \dots & 20050825 & $< 264$ \\ 
 3125 & 20050816 & $< 414$ \\ 
 3193B & 20050809 & $< 198$ \\ 
 3304 & 20050809 & $< 339$ \\ 
 \dots & 20050824 & $< 249$ \\ 
 3820 & 20050824 & $< 207$ \\ 
 4053 & 20050816 & $< 159$ \\ 
 \dots & 20050822 & $< 150$ \\ 
 4274 & 20050824 & $< 198$ \\ 
\enddata
\end{deluxetable}

\begin{deluxetable}{llrrr}
\tablecaption{VLBA Observations of M dwarfs
\label{tab:dates}}
\tablehead{
\colhead{GJ} & \colhead{Date} & \colhead{Beam Sizes} & \colhead{Beam PA} & \colhead{Image RMS} \\
               &                &  \colhead{(mas, mas)} & \colhead{(deg)} & \colhead{$\mu$Jy} }
\startdata
65B & 23 MAR 2006 A & (  3.5,  1.5) & -10.7 & 163 \\ 
65B & 23 MAR 2006 B & (  4.2,  1.8) &  11.4 & 141 \\ 
65B & 25 MAR 2006 A & (  3.4,  1.4) & -15.7 & 157 \\ 
65B & 25 MAR 2006 B & (  2.8,  2.0) &   8.4 & 118 \\ 
65B & 26 MAR 2006 A & (  3.5,  1.5) & -13.6 & 147 \\ 
65B & 26 MAR 2006 B & (  2.7,  2.1) &   9.1 & 121 \\ 
285 & 21 MAR 2006 & (  2.1,  0.9) &  -1.1 &  90 \\ 
285 & 24 MAR 2006 & (  2.5,  1.0) &  13.4 & 109 \\ 
285 & 27 MAR 2006 & (  2.7,  1.1) &  15.9 & 111 \\ 
412B & 25 MAR 2006 & (  2.4,  0.9) & -11.2 & 109 \\ 
412B & 30 MAR 2006 & (  2.2,  1.2) & -12.7 &  86 \\ 
412B & 01 APR 2006 & (  2.2,  0.9) & -12.8 &  86 \\ 
803 & 10 MAY 2006 & (  3.2,  1.9) &   5.3 & 111 \\ 
803 & 20 MAY 2006 & (  3.2,  1.6) &   6.0 & 111 \\ 
803 & 21 MAY 2006 & (  3.3,  1.7) &   8.2 & 109 \\ 
896A & 23 MAR 2006 & (  2.6,  1.5) &   2.3 &  89 \\ 
896A & 25 MAR 2006 & (  2.4,  1.0) & -10.9 &  85 \\ 
896A & 26 MAR 2006 & (  2.4,  1.0) & -10.2 &  88 \\ 
1224 & 10 MAY 2006 & (  2.9,  2.0) &  12.3 &  88 \\ 
1224 & 20 MAY 2006 & (  3.0,  1.6) &  -0.5 & 104 \\ 
1224 & 21 MAY 2006 & (  2.9,  1.8) &   0.6 & 102 \\ 
4247 & 23 MAR 2006 & (  2.4,  1.6) &  -5.2 &  95 \\ 
4247 & 25 MAR 2006 & (  2.1,  1.1) & -16.6 &  85 \\ 
4247 & 26 MAR 2006 & (  2.2,  1.1) & -16.2 &  87 \\ 
\enddata
\end{deluxetable}

\begin{deluxetable}{rrrllr}
\rotate
\tablecaption{Positions of Astrometric Calibrators \label{tab:cals}}
\tablehead{
\colhead{GJ} & \colhead{Calibrator} &  \colhead{Flux}  & \colhead{$\alpha$} & \colhead{$\delta$} &  Type \\
             &                      &  \colhead{(mJy)} &                    &                   &          
}
\startdata
65B & J0132-1654 & 1294.0 $\pm$    5.7 & 01 32 43.487468  & -16  54  48.522043  & primary \\ 
\dots & J0142-1633 &   38.4 $\pm$    1.2 & 01 42 45.261198 $\pm$ 00.000014 & -16  33  07.196786 $\pm$ 00.000121 & secondary \\ 
285 & J0739+0137 &  463.6 $\pm$    2.2 & 07 39 18.033892  & 01  37  04.617926  & primary \\ 
\dots & J0751+0152 &   50.4 $\pm$    1.4 & 07 51 02.281590 $\pm$ 00.000002 & 01  52  15.761011 $\pm$ 00.000023 & secondary \\ 
412B& J1108+4330 &  368.6 $\pm$    1.6 & 11 08 23.476932  & 43  30  53.657076  & primary \\ 
\dots & J1110+4403 &  103.7 $\pm$    2.6 & 11 10 46.345812 $\pm$ 00.000003 & 44  03  25.925528 $\pm$ 00.000028 & secondary \\ 
803 & J2056-3208 &  594.5 $\pm$    4.4 & 20 56 25.070234  & -32  08  47.800507  & primary \\ 
\dots & J2042-3152 &   62.6 $\pm$    0.6 & 20 42 46.406548 $\pm$ 00.000075 & -31  52  28.592351 $\pm$ 00.002223 & secondary \\ 
896A & J2328+1929 &  119.4 $\pm$    0.8 & 23 28 24.874755  & 19  29  58.030041  & primary \\ 
\dots & J2334+2010 &   62.2 $\pm$    1.9 & 23 34 14.156496 $\pm$ 00.000009 & 20  10  28.882640 $\pm$ 00.000279 & secondary \\ 
1224 & J1753-1843 &  110.4 $\pm$    0.5 & 17 53 09.088754  & -18  43  38.523184  & primary \\ 
\dots & J1825-1718 &   69.1 $\pm$    0.3 & 18 25 36.532398 $\pm$ 00.000029 & -17  18  49.849551 $\pm$ 00.002606 & secondary \\ 
\dots & J1809-1520 &   16.4 $\pm$    0.2 & 18 09 10.208891 $\pm$ 00.001145 & -15  20  09.692410 $\pm$ 00.030386 & secondary \\ 
4247 & J2205+2926 &  141.5 $\pm$    0.6 & 22 05 46.506426  & 29  26  55.131163  & primary \\ 
\dots & J2203+2811 &   56.8 $\pm$    1.5 & 22 03 59.147970 $\pm$ 00.000004 & 28  11  21.869208 $\pm$ 00.000015 & secondary \\ 
\enddata
\end{deluxetable}

\begin{deluxetable}{rrrrrrrrr}
\rotate
\tablewidth \textheight
\tabletypesize{\scriptsize}
\tablecaption{Positions of Stars
\label{tab:positions}}
\tablehead{
\colhead{GJ} &  \colhead{MJD} & Flux (mJy) & \colhead{RA} & \colhead{Dec.} & \colhead{$\Delta\alpha$} & \colhead{$\Delta\delta$} & \colhead{$\Delta\alpha_{{\rm opt}}$} & \colhead{$\Delta\delta_{{\rm opt}}$} \\
             &                &            & \colhead{(J2000)}      & \colhead{(J2000)}        & \colhead{(mas) }& \colhead{(mas)}& \colhead{(mas) }& \colhead{(mas)}
}
\startdata
     65B & 53817.78 & 2.171 $\pm$ 0.482 & 01 39 02.990153 $\pm$ 0.000014 & -17  56  57.918175 $\pm$ 00.000342 &   0.000 $\pm$   0.198 &   0.000 $\pm$   0.342 & 319.940 $\pm$ 155.000 & 264.745 $\pm$ 155.000  \\ 
   \dots & 53817.89 & 1.800 $\pm$ 0.340 & 01 39 02.990308 $\pm$ 0.000011 & -17  56  57.918094 $\pm$ 00.000287 &   2.208 $\pm$   0.161 &   0.081 $\pm$   0.287 & 320.487 $\pm$ 155.000 & 264.289 $\pm$ 155.000  \\ 
   \dots & 53819.89 & 1.872 $\pm$ 0.289 & 01 39 02.992267 $\pm$ 0.000008 & -17  56  57.906928 $\pm$ 00.000202 &  30.158 $\pm$   0.116 &  11.247 $\pm$   0.202 & 319.378 $\pm$ 155.000 & 266.167 $\pm$ 155.000  \\ 
   \dots & 53820.89 & 2.075 $\pm$ 0.443 & 01 39 02.993345 $\pm$ 0.000016 & -17  56  57.902321 $\pm$ 00.000456 &  45.551 $\pm$   0.229 &  15.854 $\pm$   0.456 & 320.182 $\pm$ 155.000 & 266.183 $\pm$ 155.001  \\ 
\hline
     285 & 53815.14 & 1.476 $\pm$ 0.275 & 07 44 40.017980 $\pm$ 0.000006 & 03  33  6.109089 $\pm$ 00.000163 &   0.000 $\pm$   0.088 &   0.000 $\pm$   0.163 & -45.984 $\pm$  41.247 &  73.853 $\pm$  35.409  \\ 
   \dots & 53818.14 & 0.553 $\pm$ 0.100 & 07 44 40.017550 $\pm$ 0.000007 & 03  33  6.108662 $\pm$ 00.000208 &  -6.436 $\pm$   0.112 &  -0.427 $\pm$   0.208 & -46.170 $\pm$  41.247 &  74.115 $\pm$  35.409  \\ 
\hline
     803 & 53865.20 & 1.772 $\pm$ 0.310 & 20 45 09.679701 $\pm$ 0.000011 & -31  20  29.507331 $\pm$ 00.000290 &   0.000 $\pm$   0.135 &   0.000 $\pm$   0.290 &  17.516 $\pm$  26.115 &  -3.153 $\pm$  19.137  \\ 
\hline
    896A & 53817.81 & 0.929 $\pm$ 0.203 & 23 31 52.433685 $\pm$ 0.000008 & 19  56  13.712351 $\pm$ 00.000171 &   0.000 $\pm$   0.112 &   0.000 $\pm$   0.171 & 114.239 $\pm$  35.428 &  20.565 $\pm$  28.187  \\ 
   \dots & 53819.81 & 1.399 $\pm$ 0.298 & 23 31 52.434263 $\pm$ 0.000013 & 19  56  13.714736 $\pm$ 00.000202 &   8.148 $\pm$   0.183 &   2.385 $\pm$   0.202 & 114.366 $\pm$  35.428 &  20.924 $\pm$  28.187  \\ 
   \dots & 53820.81 & 3.895 $\pm$ 0.230 & 23 31 52.434540 $\pm$ 0.000002 & 19  56  13.715419 $\pm$ 00.000045 &  12.059 $\pm$   0.031 &   3.068 $\pm$   0.045 & 114.281 $\pm$  35.428 &  20.572 $\pm$  28.186  \\ 
\hline
    4247 & 53817.79 & 0.950 $\pm$ 0.201 & 22 01 13.304385 $\pm$ 0.000008 & 28  18  25.130036 $\pm$ 00.000173 &   0.000 $\pm$   0.104 &   0.000 $\pm$   0.173 &   7.147 $\pm$ 122.891 &  85.510 $\pm$  90.706  \\ 
   \dots & 53819.79 & 1.161 $\pm$ 0.300 & 22 01 13.304782 $\pm$ 0.000016 & 28  18  25.132249 $\pm$ 00.000257 &   5.247 $\pm$   0.207 &   2.213 $\pm$   0.257 &   7.392 $\pm$ 122.892 &  85.268 $\pm$  90.706  \\ 
   \dots & 53820.79 & 1.164 $\pm$ 0.257 & 22 01 13.304933 $\pm$ 0.000011 & 28  18  25.134030 $\pm$ 00.000196 &   7.242 $\pm$   0.148 &   3.994 $\pm$   0.196 &   6.914 $\pm$ 122.891 &  85.804 $\pm$  90.706  \\ 
\enddata
\end{deluxetable}

\begin{deluxetable}{rrrrr}
\tablecaption{Proper Motion and Acceleration Relative to Optical Astrometry
\label{tab:acc}}
\tablehead{
\colhead{GJ} & \colhead{$\Delta\mu_\alpha$}& \colhead{$\Delta\mu_\delta$} &
\colhead{$a_\alpha$} & \colhead{$a_\delta$} \\
            & \colhead{(mas${\rm\, y^{-1}}$)}           & \colhead{(mas${\rm\, y^{-1}}$)}           & \colhead{(AU ${\rm\, y^{-2}}$)}           & \colhead{(AU ${\rm\, y^{-2}}$)}           
}
\startdata
     65B &  37.3 $\pm$  52.5 & -218.5 $\pm$  38.8 &  0.0032 $\pm$  0.0045 & -0.0189 $\pm$  0.0033 \\ 
     285 &  22.4 $\pm$  12.5 & -32.1 $\pm$  22.7 &  0.0088 $\pm$  0.0049 & -0.0127 $\pm$  0.0090 \\ 
    896A &  -7.4 $\pm$   7.2 & -10.1 $\pm$  24.5 & -0.0031 $\pm$  0.0030 & -0.0042 $\pm$  0.0102 \\ 
    4247 &  18.1 $\pm$  27.8 & -24.4 $\pm$  30.3 &  0.0046 $\pm$  0.0071 & -0.0062 $\pm$  0.0078 \\ 
\enddata
\end{deluxetable}

\begin{deluxetable}{lrrrr}
\tablecaption{Limits in Companion Mass and Semi-Major Axis
\label{tab:limits}}
\tablehead{\colhead{GJ} & \colhead{$r_{min}$} & \colhead{$M_{p,min}$} & \colhead{$M_{p,0.3}$} & \colhead{$M_{p,3}$} \\
& \colhead{(AU)} & \colhead{($M_J$)}& \colhead{($M_J$)}& \colhead{($M_J$)}}
\startdata
65B & 1.3 & 4.5 & 19 & 24\\
285 & 0.8 & 3.3 & 9 & 42\\
896A & 0.8 & 3.8 & 10 & 42 \\
4247 & 1.0 & 6.1 & 21 & 44 \\
\enddata
\end{deluxetable}
\end{document}